\documentclass[preprintnumbers,amssymb,amsmath,superscriptaddress,letterpaper,nofootinbib,twocolumn,floatfix]{revtex4}[10pt]

\usepackage{graphicx}
\usepackage{dcolumn}
\usepackage{bm}
\usepackage{natbib}
\usepackage{calligra}
\usepackage[T1]{fontenc}
\usepackage{egothic}
\usepackage[T1]{fontenc}
\newfont{\rsfsten}{rsfs10 scaled 1200}
\newfont{\rsfsseven}{rsfs10 scaled 1200}
\newfont{\rsfsfive}{rsfs10 scaled 1200}
\usepackage{epsfig}
\usepackage{units}

\newcommand{\be}{\begin{equation}}
\newcommand{\ee}{\end{equation}}
\newcommand{\bea}{\begin{eqnarray}}
\newcommand{\eea}{\end{eqnarray}}

\def\lsim{\mathrel{\raise.3ex\hbox{$<$\kern-.75em\lower1ex\hbox{$\sim$}}}}
\def\gsim{\mathrel{\raise.3ex\hbox{$>$\kern-.75em\lower1ex\hbox{$\sim$}}}}

\begin{document}

\hspace*{130mm}{\large \tt FERMILAB-PUB-15-559-A}
\vskip 0.2in

\title{The Gamma-Ray Luminosity Function of Millisecond Pulsars and Implications for the GeV Excess}


\author{Dan Hooper}
\affiliation{Fermi National Accelerator Laboratory, Center for Particle Astrophysics, Batavia, IL}
\affiliation{University of Chicago, Department of Astronomy and Astrophysics, Chicago, IL}
\author{Gopolang Mohlabeng}
\affiliation{Fermi National Accelerator Laboratory, Center for Particle Astrophysics, Batavia, IL}
\affiliation{University of Kansas, Department of Physics and Astronomy, Lawrence, KS}
\date{\today}

\begin{abstract}

It has been proposed that a large population of unresolved millisecond pulsars (MSPs) could potentially account for the excess of GeV-scale gamma-rays observed from the region surrounding the Galactic Center. The viability of this scenario depends critically on the gamma-ray luminosity function of this source population, which determines how many MSPs Fermi should have already detected as resolved point sources. In this paper, we revisit the gamma-ray luminosity function of MSPs, without relying on uncertain distance measurements. Our determination, based on a comparison of models with the observed characteristics of the MSP population, suggests that Fermi should have already detected a significant number of sources associated with such a hypothesized Inner Galaxy population. We cannot rule out a scenario in which the MSPs residing near the Galactic Center are systematically less luminous than those present in the Galactic Plane or within globular clusters.

\end{abstract}


\maketitle

\section{Introduction}

As a result of the ongoing operation of the Fermi Gamma-Ray Space Telescope, our knowledge of gamma-ray pulsars has dramatically expanded.  To date, Fermi has detected more than 160 pulsars, establishing them as the dominant class of GeV-scale sources in the Milky Way~\cite{2009Sci...325..848A,TheFermi-LAT:2013ssa}.  Particularly pronounced has been the progress made regarding pulsars with millisecond-scale periods, generally believed to be old neutron stars that were spun-up to extremely rapid rates of rotation through the transfer of angular momentum from a companion star~\cite{1982Natur.300..728A,1994ARA&A..32..591P,Lorimer:2001vd,Lorimer:2008se,2010ApJ...715..335K}. In contrast to young pulsars, millisecond pulsars (MSPs) have weaker magnetic fields and thus lose their rotational kinetic energy more slowly, typically remaining luminous for billions of years.  Fermi has detected emission from 70 MSPs, none of which were previously observed at gamma-ray wavelengths~\cite{TheFermi-LAT:2013ssa,Guillemot:2011th,2009Sci...325..848A}.\footnote{For a list of Fermi-detected pulsars, see \url{https://confluence.slac.stanford.edu/display/GLAMCOG/Public+List+of+LAT-Detected+Gamma-Ray+Pulsars}.}

The long lifetimes of MSPs have lead some to speculate that large numbers of such objects could exist in and around the Milky Way's disk, plausibly contributing significantly to the diffuse gamma-ray background~\cite{FaucherGiguere:2009df} (for early work in this direction, see Ref.~\cite{1996ApJ...461..872S}). The characteristics of the MSP population detected by Fermi, however, reveal that luminous MSPs in the field of the Milky Way are too rare to generate a significant fraction of the diffuse gamma-ray background~\cite{Hooper:2013nhl,Gregoire:2013yta,Calore:2014oga} (see also, Ref.~\cite{SiegalGaskins:2010mp}). 

A bright excess of GeV-scale gamma-rays from the region surrounding the Galactic Center has been identified from within the publicly available Fermi dataset~\cite{Goodenough:2009gk,Hooper:2010mq,Hooper:2011ti,Abazajian:2012pn,Gordon:2013vta,Hooper:2013rwa,Daylan:2014rsa,Calore:2014xka}, and recently confirmed by the Fermi Collaboration~\cite{TheFermi-LAT:2015kwa}. This signal has a spectrum, angular distribution, and overall normalization that is consistent with the expectations from annihilating dark matter, stimulating a great deal of interest (see, for example, Refs.~\cite{Abdullah:2014lla,Ipek:2014gua,Izaguirre:2014vva,Agrawal:2014una,Berlin:2014tja,Alves:2014yha,Boehm:2014hva,Martin:2014sxa,Huang:2014cla,Cerdeno:2014cda,Okada:2013bna,Freese:2015ysa,Fonseca:2015rwa,Bertone:2015tza,Cline:2015qha,Berlin:2015wwa,Caron:2015wda,Cerdeno:2015ega,Liu:2014cma,Hooper:2014fda,Arcadi:2014lta,Cahill-Rowley:2014ora,Ko:2014loa,McDermott:2014rqa,Kong:2014haa}). The leading astrophysical interpretation of this signal is that it is generated by a large population of unresolved MSPs, in a highly concentrated and approximately spherical distribution around the Galactic Center~\cite{Hooper:2010mq,Hooper:2011ti,Abazajian:2012pn,Gordon:2013vta,Yuan:2014rca,Petrovic:2014xra,Brandt:2015ula}. The primary motivation for this possibility is the spectral shape of the GeV excess, which is similar to that observed from MSPs. Spurring further attention on this scenario are the results of two recent studies, which find that the gamma-rays from the region of the excess are more spatially clustered than statistically expected~\cite{Lee:2015fea,Bartels:2015aea}. At this time, however, it is not clear whether these techniques are detecting the presence of a sub-threshold point source population or merely groups of photons associated with the small scale structure of the diffuse background. We are hopeful that these and other~\cite{McDermott:2015ydv,probcat} analysis strategies will help to clarify this situation in the near future. 

Previous studies have concluded that if the GeV excess does, in fact, originate from MSPs, then the inner kiloparsecs of the Milky Way should contain many more bright MSPs than have been detected by Fermi~\cite{Cholis:2014lta,SiegalGaskins:2010mp}. From these results, it appears that MSPs could account for the GeV excess only if the luminosity function of this pulsar population was significantly different (containing a smaller fraction of high-luminosity members)~\cite{Yuan:2014rca,Petrovic:2014xra} than those populations observed in the field of the Milky Way or within globular clusters~\cite{Cholis:2014noa}.  It has been suggested, however, that the determination of the luminosity function presented in Ref.~\cite{Cholis:2014noa} (and utilized in Ref.~\cite{Cholis:2014lta}) may be inaccurate due to the systematic mismeasurement of the distances to many MSPs, perhaps allowing for the possibility that Fermi might not be sensitive to many of the MSPs responsible for the GeV excess~\cite{Brandt:2015ula}.\footnote{For most of the MSPs observed by Fermi, distances have only been estimated from radio dispersion measurements. As this association relies on models of the interstellar electron distribution, such determinations can involve significant and difficult to quantify uncertainties~\cite{TheFermi-LAT:2013ssa}.}



In this paper, we constrain the characteristics of the Milky Way's MSP population, while avoiding the use of uncertain distance determinations based on radio dispersion measurements. The remainder of this paper is structured as follows. In Sec.~\ref{modeling}, we construct a model for the spatial distribution and luminosity function of the MSP population associated with the Milky Way's disk, and constrain this population with the angular and flux distribution observed by Fermi, taking care to accurately model Fermi's direction- and luminosity-dependent MSP detection probability. In Sec.~\ref{periods}, we consider measurements of MSP periods and their time-derivatives within the context of the luminosity function inferred from our fits. In Sec.~\ref{lumfuncsec}, we present our determination of the MSP luminosity function, taking into account the results of our model fit, as well as the characteristics of the gamma-ray emission observed from the Milky Way's globular cluster population.  In Sec.~\ref{innermw}, we utilize our derived luminosity function to determine the number of MSPs that should have been detected by Fermi if such sources are responsible for the GeV excess, finding quantities that exceed the number of observed MSP candidates. Finally, we summarize our results and conclusions in Sec.~\ref{conclusions}. 

\section{Modeling The Milky Way's Millisecond Pulsar Population}
\label{modeling}

To model and ultimately constrain the Milky Way's MSP population, we have utilized a Monte Carlo which draws from a distribution for each pulsar's location and gamma-ray luminosity.  We take the luminosity function to be described by a log-normal distribution, who's central value and width, $L_0$ and $\sigma_L$, are treated as free parameters.\footnote{This is in contrast to the power-law or truncated power-law parameterizations that have been adopted in other recent studies~\cite{Petrovic:2014xra,Bartels:2015aea}.} The gamma-ray luminosity of a given MSP is expected to be a function of its magnetic field, rotational period, and gamma-ray efficiency. In Section~\ref{periods}, we will discuss the luminosity function within the context of these characteristics of the underlying MSP population. For the spatial distribution of MSPs associated with the Milky Way's disk (as opposed to those in globular clusters~\cite{Fermi:2011bm,collaboration:2010bb,msps_47Tuc, msps_10_47Tuc, msps_more_47Tuc, Camilo:1999fc,Bogdanov:2006ap} or in the Milky Way's bulge or central stellar cluster~\cite{Brandt:2015ula}), we adopt the following parameterization:
\begin{equation}
\frac{dn_{\rm MSP}}{dV} \propto e^{-|z|/z_{0}} \, e^{-r^2/2 \sigma_R^2},
\end{equation}
where $r$ and $z$ describe the location of a pulsar in cylindrical coordinates, and $z_0$ and $\sigma_R$ are free parameters. 


In order to compare the MSP population generated by our Monte Carlo with the population observed and reported by the Fermi Collaboration, we must first establish the probability that a given pulsar will be detected by Fermi.  In particular, Fermi's detection threshold varies significantly over the sky, being significantly higher in regions with large backgrounds, such as along the Galactic Plane. To address this, we adopt a direction-dependent detection threshold that is proportional to the Fermi pulsar flux sensitivity, as shown in Fig.~16 of Ref.~\cite{TheFermi-LAT:2013ssa}.\footnote{We thank Matthew Kerr and David Smith for providing us with the numerical values that constitute this map.} The constant of proportionality is allowed to float and is treated as a nuisance parameter in our fits. We additionally vary the detection threshold on a pulsar-to-pulsar basis around the central value according to a log-normal distribution with a width of $\sigma=0.9$. This width was selected to achieve approximate agreement with the distribution shown in Fig.~17 of Ref.~\cite{TheFermi-LAT:2013ssa}, and with the observed near-threshold flux distribution. We have checked that modest variations of this quantity do not qualitatively alter our conclusions.

\begin{figure*}
\includegraphics[width=3.0in,angle=0]{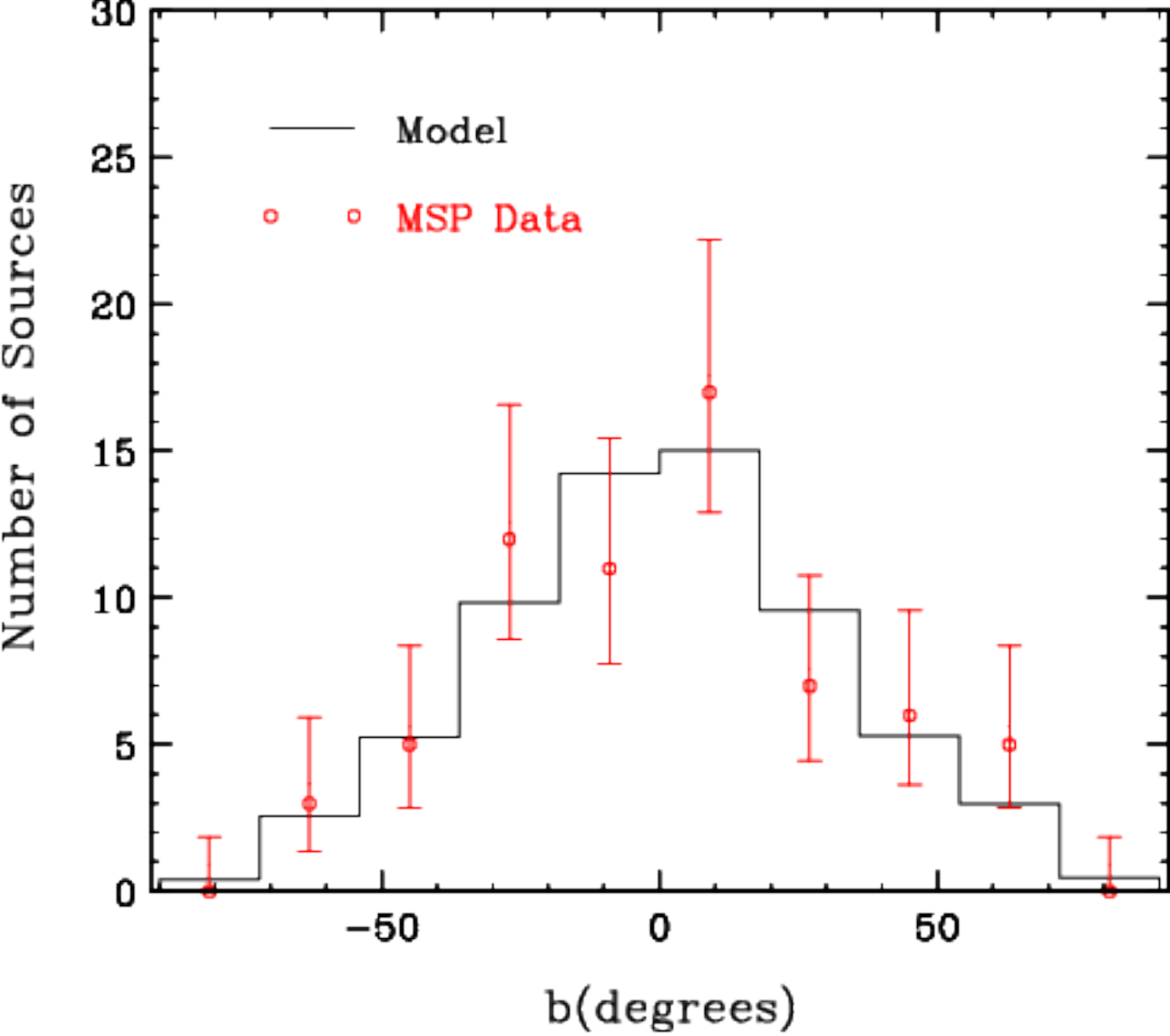} \hspace{0.7cm}
\includegraphics[width=3.0in,angle=0]{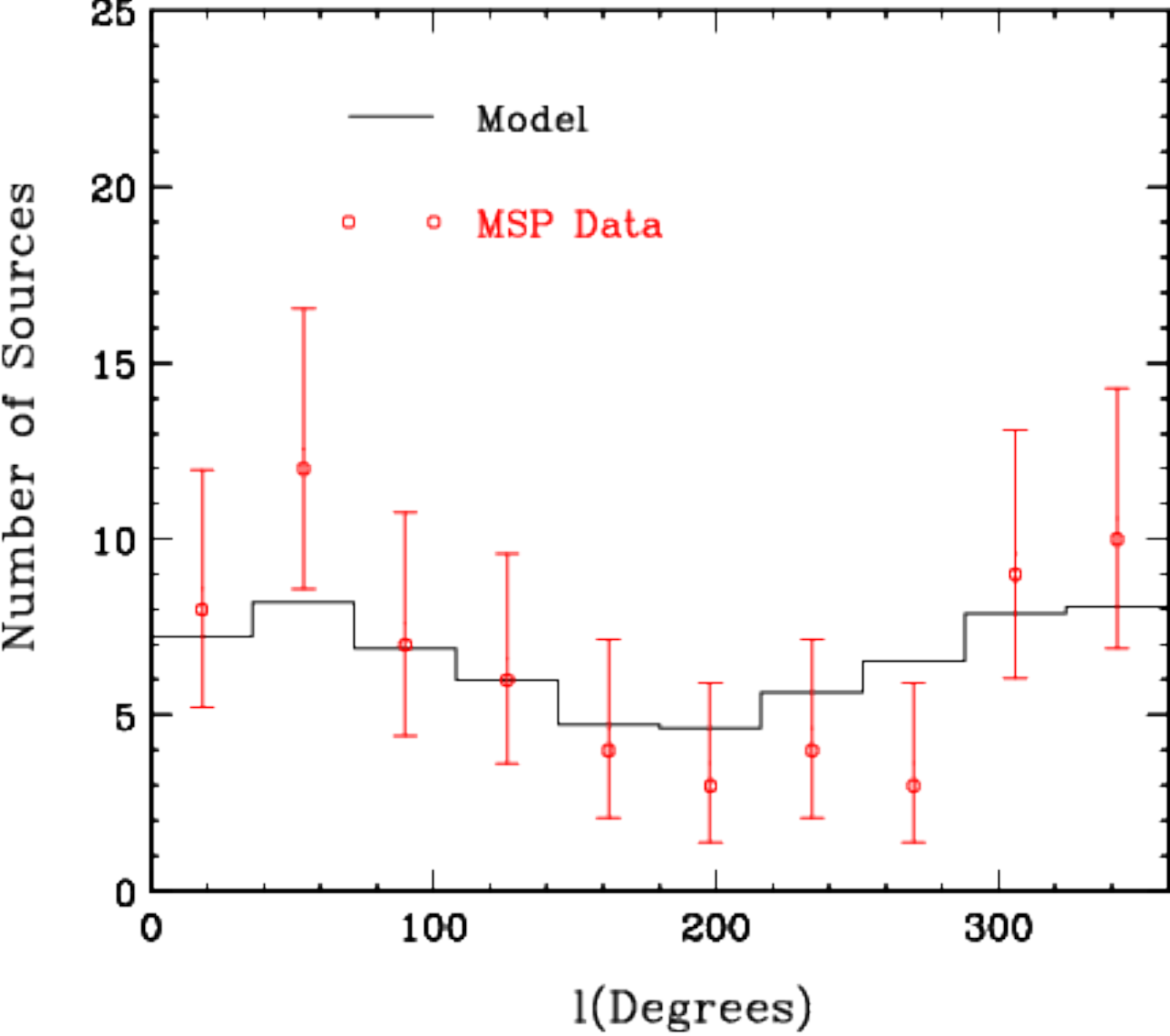} \\
\vspace{0.5cm}
\includegraphics[width=3.0in,angle=0]{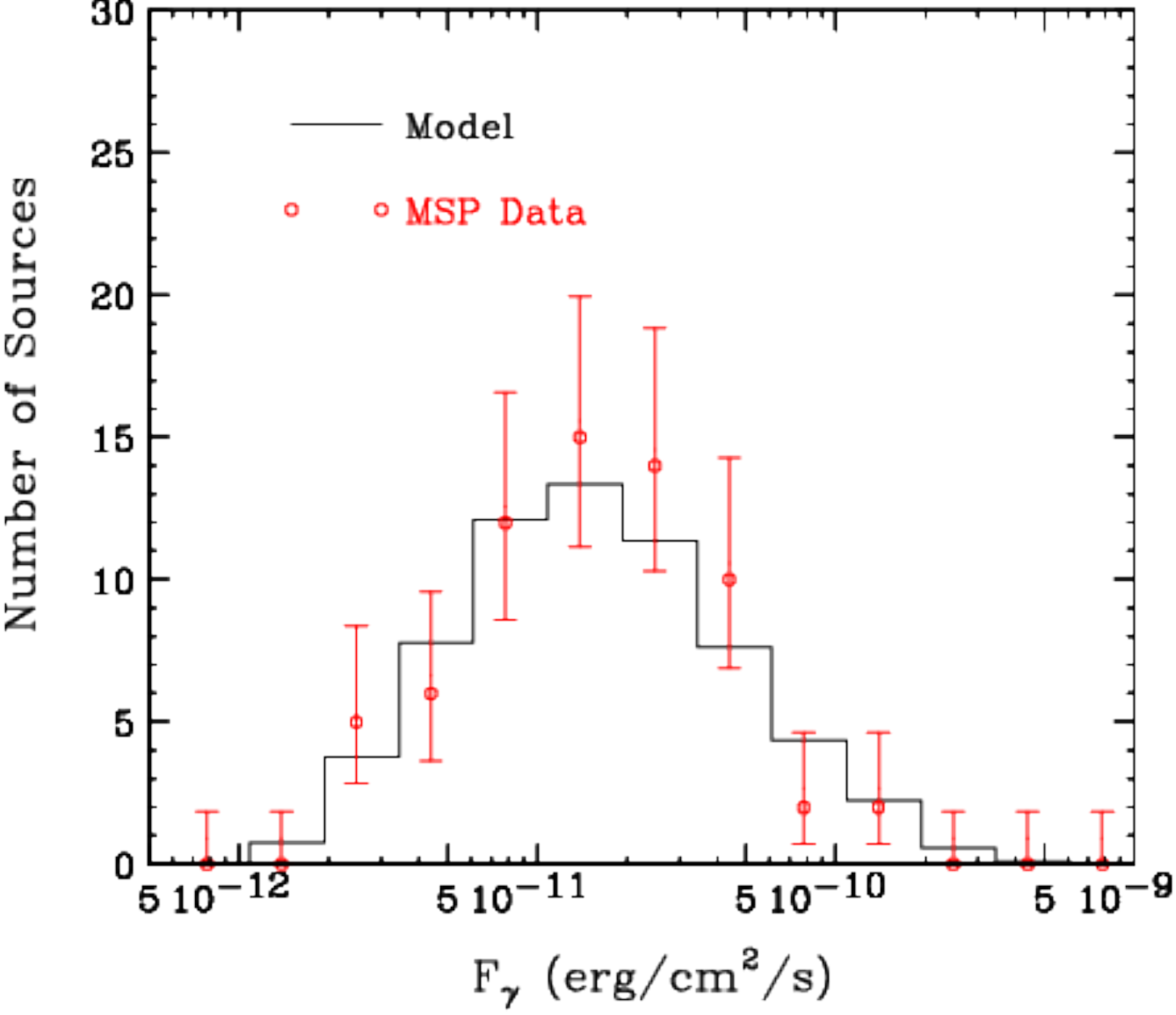} 
\caption{The distribution in Galactic Latitude, Galactic Longitude, and gamma-ray flux (integrated above 0.1 GeV) of the millisecond pulsars detectable by Fermi as predicted by our best-fit model ($z_0=0.39$ pc, $\sigma_R=4.2$ kpc, $L_0=4.7 \times 10^{32}$ erg/s, and $\sigma_L=1.4$).  This is compared to the distribution observed by Fermi (red error bars). This model provides a good statistical fit to the data.}
\label{histograms}
\end{figure*}

\begin{figure*}
\includegraphics[width=3.40in,angle=0]{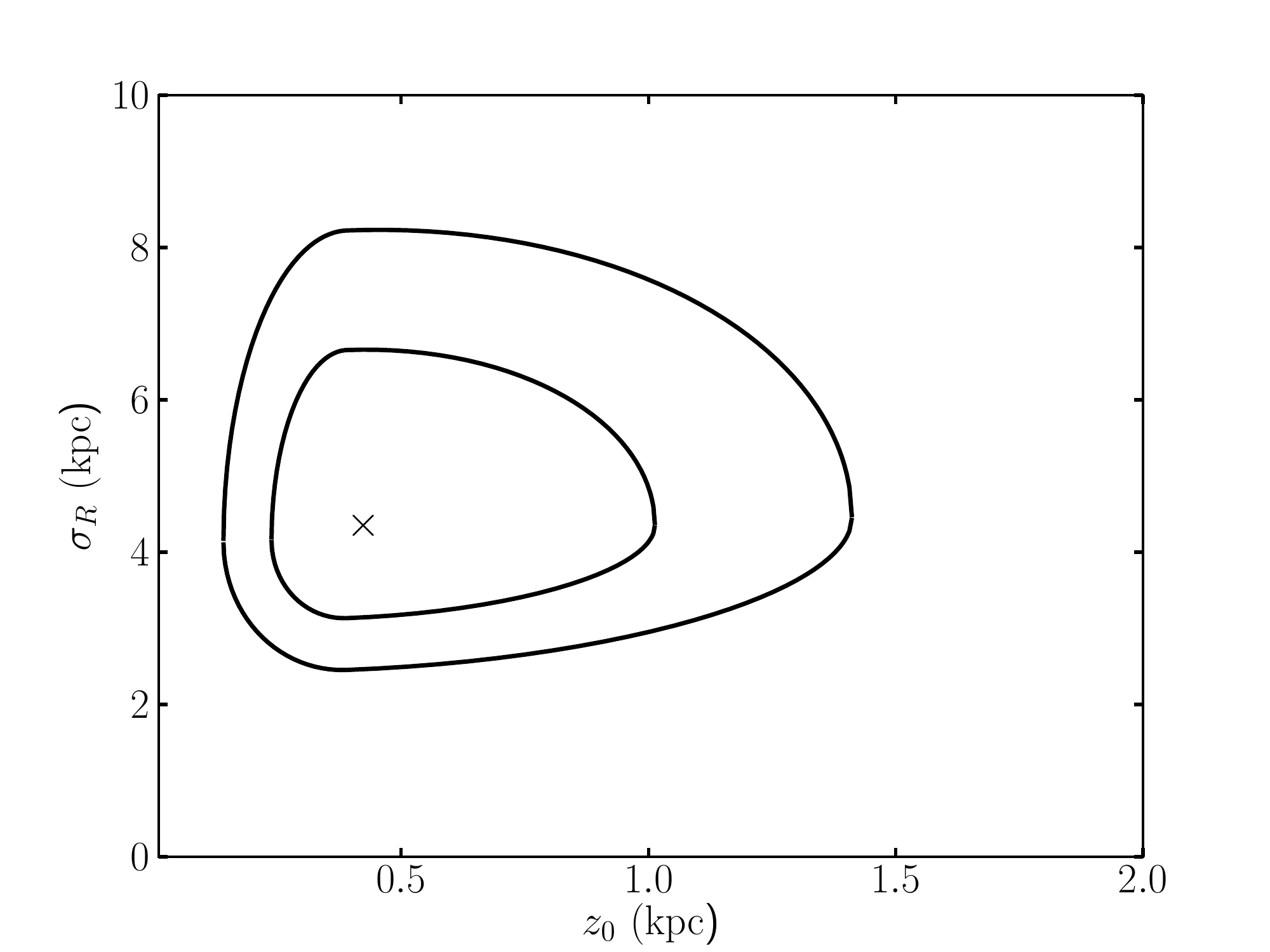} 
\includegraphics[width=3.40in,angle=0]{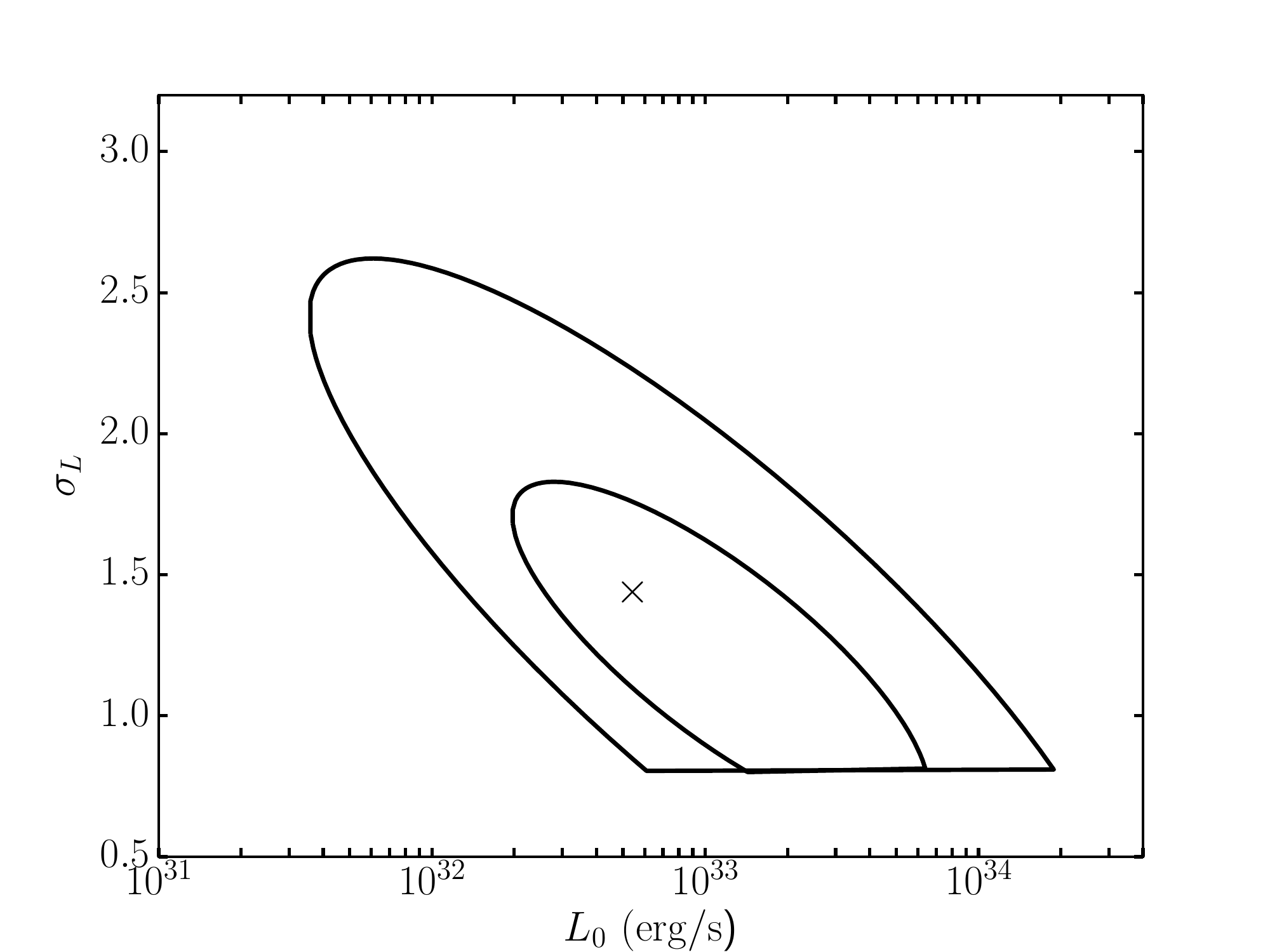} 
\caption{The determination of the parameters describing the spatial distribution (left) and gamma-ray luminosity function (right) of the millisecond pulsar population associated with the disk of the Milky Way (as opposed to those contained within globular clusters, or within our galaxy's bulge or central stellar cluster). The crosses denote the best-fit point in each plane. Pulsar distance estimates, which may suffer from significant and difficult to quantify uncertainties, have not been used in this determination.}
\label{ellipses}
\end{figure*}

For each choice of parameters, we compare the output of our Monte Carlo with the latitude, longitude, and flux distribution of MSPs observed by Fermi. In this comparison, we include the 66 gamma-ray detected pulsars found within the Third Fermi Source Catalog (3FGL)~\cite{Acero:2015hja} that exhibit rotational periods less than 10 milliseconds, as reported in the Australia Telescope National Facility (ATNF) pulsar catalog~\cite{Manchester:2004bp}. From this comparison with the data, we calculate the likelihood for each parameter set ($z_0$, $\sigma_R$, $L_0$, $\sigma_L$):
\begin{equation}
\mathcal{L} = \prod_i \frac{M_i^{D_i} e^{-M_i}}{D_i!},
\end{equation}
where $M_i$ and $D_i$ respectively denote the numbers of MSPs predicted by our Monte Carlo and observed by Fermi within the $i$th bin. In each case, we normalize our model such that 66 MSPs are observed by Fermi. 
The best fit was found for the case $z_0=0.39$ pc, $\sigma_R=4.2$ kpc, $L_0=4.7 \times 10^{32}$ erg/s ($>0.1$ GeV), and $\sigma_L=1.4$. In Fig.~\ref{histograms}, we plot these three distributions for our best-fit parameter set. This model clearly provides a good fit to the data, corresponding to $\ln {\mathcal L} \simeq -57.5$. Around this best-fit point, however, there is a significant range of parameter space that is consistent with the data. In Fig.~\ref{ellipses}, we plot the 1$\sigma$ and 2$\sigma$ contours favored by our fit in the $z_0$-$\sigma_R$ and $L_0$-$\sigma_L$ planes.\footnote{Due to computational limitations, the results of our Monte Carlo include statistical fluctuations. The contours shown in Fig.~\ref{ellipses} are smooth curves, fit to the more jagged output of our code.}  These results approximately correspond to: $z_0 = 0.4^{+0.6}_{-0.15}$ pc, $\sigma_R=4.2^{+2.4}_{-1.1}$ kpc, $\log_{10} L_0=32.7^{+1.2}_{-0.4}$, and $\sigma_L<1.8$ (each at the 1$\sigma$ level).

Although we have chosen to avoid the use of most distance measurements in this study, we do take into account those obtained by use of stellar parallax (as opposed to the less reliable determinations based on radio dispersion measurements).  Among Fermi's nine MSPs with parallax distance measurements, the gamma-ray luminosities vary between $5\times 10^{31}$ to $3\times 10^{33}$ erg/s~\cite{TheFermi-LAT:2013ssa}, which is well centered around our best-fit value of $L_0=4.7 \times 10^{32}$ erg/s. The distribution of luminosities exhibited among these nine pulsars allows us to restrict the width of the luminosity function; based on this we conservatively restrict the parameter space to those models with $\sigma_L > 0.8$.

In terms of the spatial distribution of MSPs favored by our fit, our results are similar to the results of previous studies~\cite{SiegalGaskins:2010mp,Calore:2014oga,Gregoire:2013yta}. For example, Ref.~\cite{SiegalGaskins:2010mp} found that $z_0\simeq 0.5-1$ kpc and $\sigma_R\simeq 3-7$ kpc could accommodate the observed population. Similarly, the authors of Ref.~\cite{Calore:2014oga} find $z_0=0.67 \pm 0.11$ kpc, which lies well within the 1$\sigma$ region shown in Fig.~\ref{ellipses}. We will compare our luminosity function determination with the results of previous studies in Sec.~\ref{lumfuncsec}.


\section{Rotational Periods, Magnetic Fields, and Gamma-Ray Efficiencies}
\label{periods}

The radio and gamma-ray emission observed from pulsars is powered by their rotational kinetic energy, which is transferred through the process of magnetic-dipole breaking. The rate of the decline of a pulsar's rotational period is determined by the strength of its magnetic field:
\begin{equation}
\dot{P} \simeq 3.3 \times 10^{-20} \, \bigg(\frac{B}{10^{8.5}\, {\rm G}}\bigg)^2 \, \bigg(\frac{P}{3 \, {\rm ms}}\bigg)^{-1}.
\label{pdot}
\end{equation}
This corresponds to the following rate for the loss of rotational kinetic energy:
\begin{eqnarray}
\dot{E}&=&\frac{4 \pi^2 I \dot{P}}{P^3} \\
&\simeq & 4.8 \times 10^{34} \, {\rm erg/s} \,\, \bigg(\frac{B}{10^{8.5}\, {\rm G}}\bigg)^2 \, \bigg(\frac{P}{3 \, {\rm ms}}\bigg)^{-4}, \nonumber
\end{eqnarray}
where we have taken the neutron star's moment of inertia to be $10^{45}$ g/cm$^2$. The (isotropic equivalent) gamma-ray luminosity is defined as the total energy budget, $\dot{E}$, multiplied by an efficiency factor, $L_{\gamma} \equiv \dot{E}\, \eta$. Note that a pulsar's emission can be highly anisotropic, allowing for efficiency factors (in the observed direction) that are greater than unity.

\begin{figure}
\includegraphics[width=3.40in,angle=0]{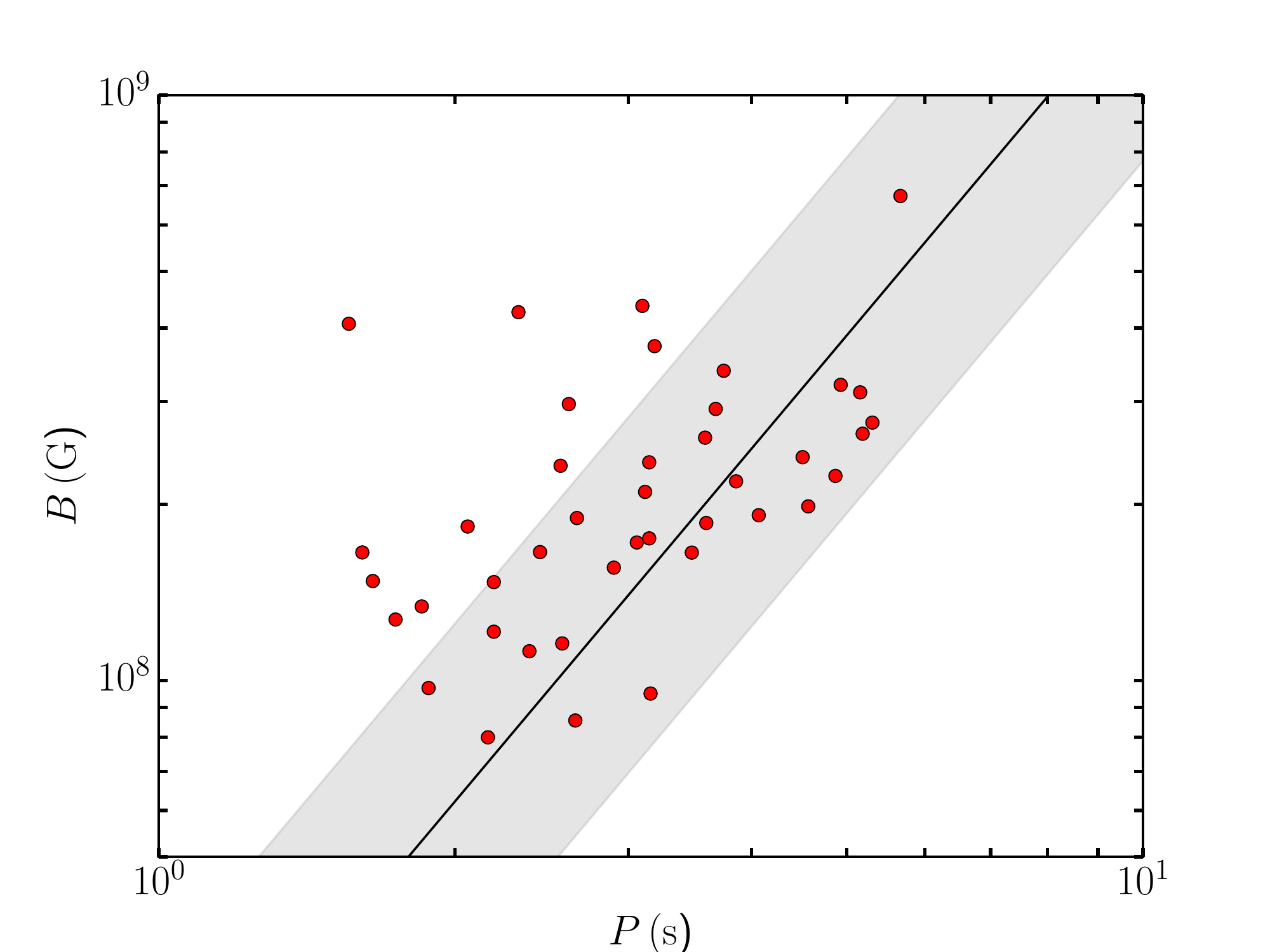} 
\caption{The observed period and magnetic field strength (as determined from $\dot{P}$ and $P$ via Eq.~\ref{pdot}), for those gamma-ray millisecond pulsars with a measured value of $\dot{P}$. Note that pulsars with large magnetic fields and/or low periods are generally more likely to be detected by Fermi, significantly biasing the distribution shown. The black line is a contour of constant luminosity, $L_{\gamma} = 4.7 \times 10^{32}$ erg/s $\times (\eta/0.05)$, surrounded by a shaded region representing the range in which 68\% of MSPs (detected or otherwise) are predicted to reside for $\sigma_L=1.4$ and neglecting any pulsar-to-pulsar variation in $\eta$.}
\label{ppdot}
\end{figure}

If we naively take the population of MSPs to have distributions of periods, magnetic fields and gamma-ray efficiencies that are described by uncorrelated log-normal distributions, we can describe the luminosity function in terms of these individual parameters: 
\begin{eqnarray}
L_0 &\simeq& 4.8 \times 10^{33} \, {\rm erg/s} \,\, \bigg(\frac{B_0}{10^{8.5}\, {\rm G}}\bigg)^2 \, \bigg(\frac{P_0}{3 \, {\rm ms}}\bigg)^{-4} \, \bigg(\frac{\eta_0}{0.1}\bigg), \nonumber \\
\sigma_L &=& 4 \sigma_P + 2 \sigma_B + \sigma_{\eta}.
\label{nocor}
\end{eqnarray}
In reality, it is not likely that these quantities are uncorrelated. The beaming geometry of a given MSP (on which values of $\eta$ rely), for example, may very well be dependent on its rotational period and/or magnetic field strength. In Fig.~\ref{ppdot}, we plot the period and magnetic field (as determined from $\dot{P}$ and $P$ via Eq.~\ref{pdot}), for each gamma-ray MSP with a measured value of $\dot{P}$. One should interpret the information presented in this figure carefully for at least two reasons. First, not all of Fermi's MSPs have a measured value of $\dot{P}$ (and thus an inferred value of $B$), and it is not clear how the selection of this subset of pulsars might bias the distribution of other parameters under consideration. Second, pulsars with large values of $B$ and/or low values of $P$ are generally brighter and thus more likely to be detected by Fermi. This explains why comparatively few MSPs are found in the lower-right portion of this figure. We also include in this frame a contour of constant luminosity, $L_{\gamma} = 4.7 \times 10^{32}$ erg/s $\times (\eta/0.05)$. The shaded region around that contour represents the range in which 68\% of MSPs (detected or otherwise) are predicted to reside, for $L_0=4.7 \times 10^{32}$ erg/s $\times (\eta/0.05)$ and $\sigma_L=1.4$, assuming (unrealistically) that any pulsar-to-pulsar variation in $\eta$ is negligible.

\begin{figure}
\includegraphics[width=3.0in,angle=0]{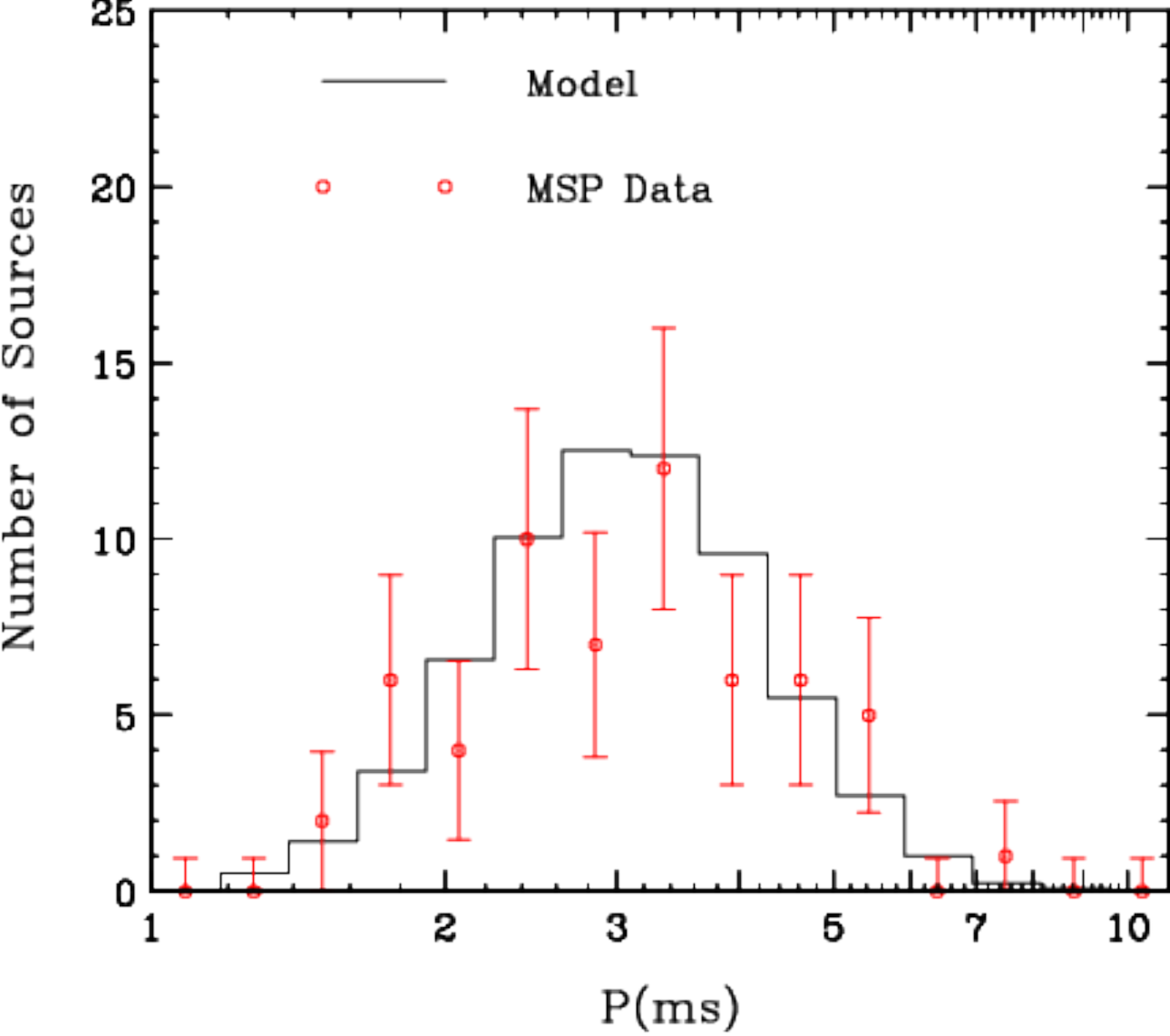} 
\caption{The distribution of the periods of the millisecond pulsars detectable by Fermi as predicted by our best-fit model ($z_0=0.39$ pc, $\sigma_R=4.2$ kpc, $L_0=4.7 \times 10^{32}$ erg/s, $\sigma_L=1.4$, $P_0=4.9$ ms, and $\sigma_p=0.35$).  This is compared to the distributions observed by Fermi (red error bars). This model provide a good statistical fit to the data.}
\label{period}
\end{figure}

\begin{figure*}
\includegraphics[width=3.40in,angle=0]{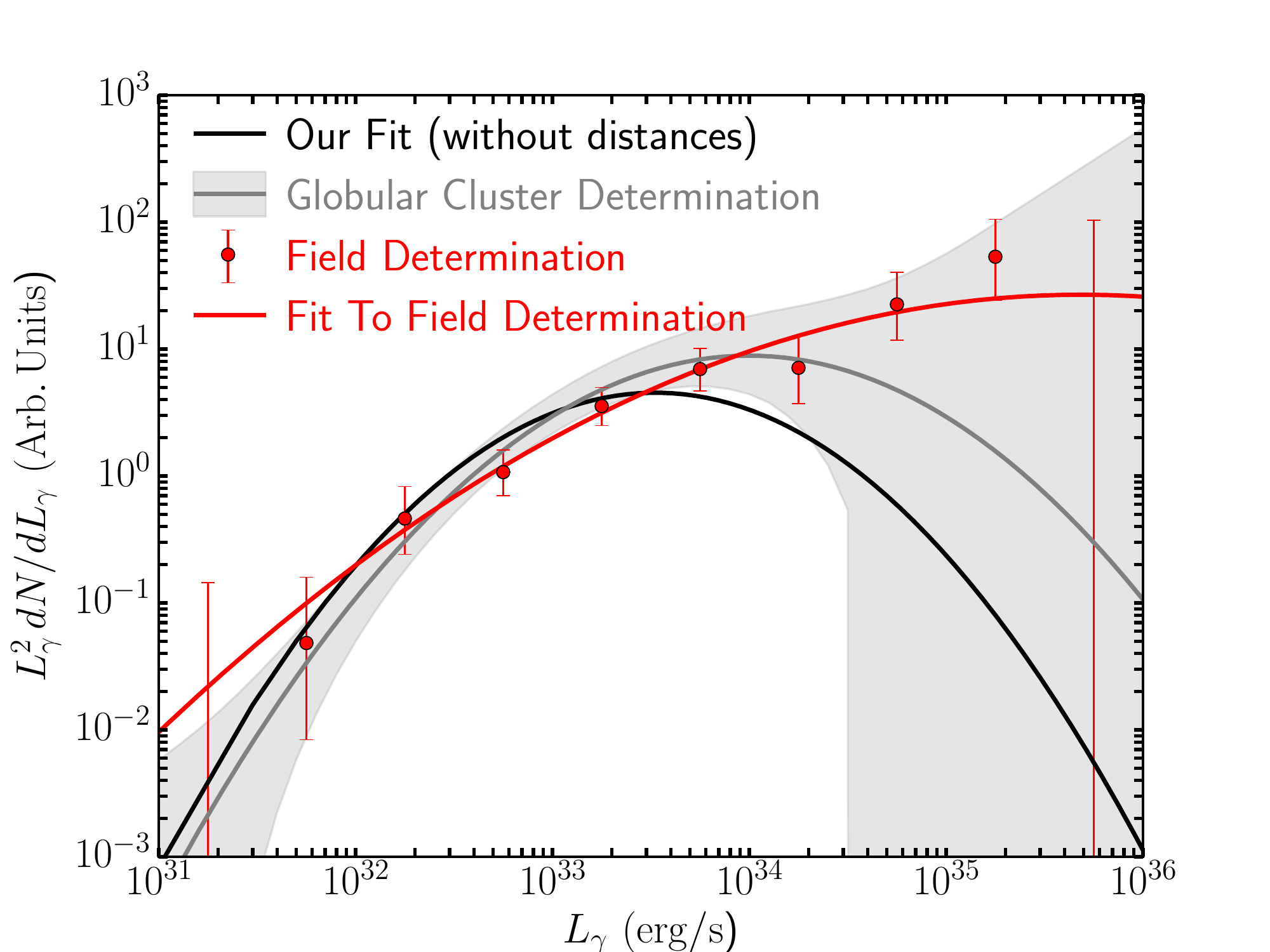} 
\includegraphics[width=3.40in,angle=0]{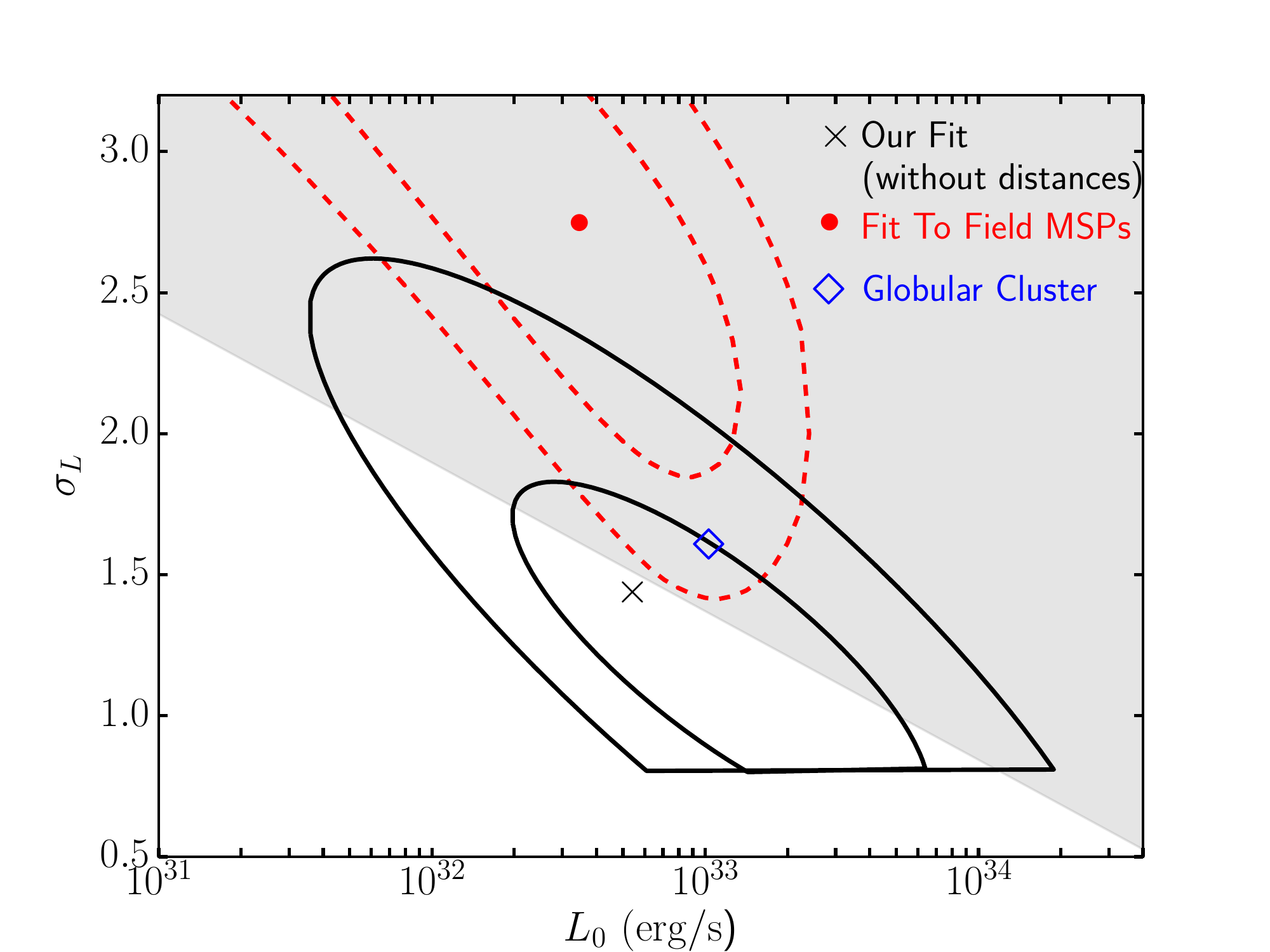} 
\caption{Left frame: the best-fit luminosity functions as derived in Sec.~\ref{modeling} of this study (black solid), from the field population of MSPs, including uncertain distance determinations~\cite{Cholis:2014noa} (red solid, and error bars), and using the MSP population within the globular cluster Tuc 47~\cite{Cholis:2014noa} (grey soild, and shaded region). Right frame: the best-fit values of $L_0$ and $\sigma_L$ for each of these three luminosity functions, as well as the 1 and 2$\sigma$ contours around the determination presented in Sec.~\ref{modeling} of this study, and the determination using field MSPs from Ref.~\cite{Cholis:2014noa}. The grey shaded region denotes the range of parameters that is consistent (at the 95\% confidence level) with the observed luminosity of PSR J1823-3021A.}
\label{lumfunc}
\end{figure*}

Any bias associated with Fermi's detection threshold can be directly addressed by our Monte Carlo. In Fig.~\ref{period}, we plot the distribution of periods for those MSPs detectable by Fermi in our best-fit model, and compare that to the observed distribution. Along with the best-fit parameters identified in the previous section ($z_0=0.39$ pc, $\sigma_R=4.2$ kpc, $L_0=4.7 \times 10^{32}$ erg/s, and $\sigma_L=1.4$), the best fit was found for $P_0 = 4.9$ ms and $\sigma_P=0.35$, similar to that found in Ref.~\cite{Lorimer:2015iga}. If we assume that there is no correlation between the period, magnetic field, and gamma-ray efficiency of a given MSP, these parameter in turn imply $B_0 \simeq 3.8 \times 10^8 \, {\rm G} \, \times (0.05/\eta_0)^{1/2}$, which is consistent with previous determinations~\cite{SiegalGaskins:2010mp,Calore:2014oga}. More problematic are the the best-fit values found for $\sigma_P$ and $\sigma_L$ which, in the absence of correlations, implies essentially no pulsar-to-pulsar variation in $B$ or $\eta$ ($\sigma_B \simeq \sigma_{\eta} \simeq 0$). As this is clearly not the case (see, for example, the range of $B$ values shown in Fig.~\ref{ppdot}), we conclude that the true MSP luminosity function is likely to feature a somewhat wider distribution than our best-fit model, likely in the range of $\sigma_L$$\sim$\,1.5-2.2. We will discuss further support for this conclusion in the following section. Additionally, this apparent problem is potentially associated in part with actual correlations that exist between these parameters. In particular, past studies have found typical behavior of the form $L_{\gamma} \propto  \dot{E}^{\alpha}$, with $\alpha \simeq 0.5 - 1.0$~\cite{1996A&AS..120C..49A,1975ApJ...196...51R,1981ApJ...245..267H,2006ApJ...643..332F,SiegalGaskins:2010mp,Hooper:2013nhl,Calore:2014oga,Gregoire:2013yta}. For $\alpha < 1$, this has the effect of reducing the value of $\sigma_L$ relative to that found if one assumes no correlations between these parameters (as in Eq.~\ref{nocor}).

\section{The MSP Gamma-Ray Luminosity Function}
\label{lumfuncsec}

Previous studies have used a number of different methods to derive the gamma-ray luminosity function of MSPs. In Ref.~\cite{Cholis:2014noa}, the authors estimated the distance to which Fermi's catalog of high-latitude ($|b|>10^{\circ}$) MSPs in the field of the Milky Way would be complete above a given luminosity, and used the distribution of such sources to infer the underlying luminosity function. The results of this approach are shown as error bars in the left frame of Fig.~\ref{lumfunc}. This determination relies on the reliability of pulsar distance measurements (Ref.~\cite{Cholis:2014noa} adopted distances as reported in the ATNF catalog~\cite{Manchester:2004bp}), however, and thus is subject to significant systematic uncertainties. The same authors also made use of MSPs observed within the globular cluster 47 Tucanae (also known as 47 Tuc, or NGC 104), combined with a previously identified empirical correlation relating their X-ray and gamma-ray luminosities~\cite{TheFermi-LAT:2013ssa}, to derive the gamma-ray luminosity function for the MSPs contained within that system~\cite{Cholis:2014noa}. Notably, this yielded results that were consistent (within uncertainties) to those obtained using the previously described method. 

In Secs.~\ref{modeling} and ~\ref{periods} of this study, we derived constraints on the MSP luminosity function by comparing the predictions of population models to the observed distribution of MSPs on the sky and in flux, without making use of uncertain distance determinations. Although this approach avoids the systematic uncertainties associated with MSP distances, it yields results which suffer from fairly large statistical uncertainties, as can be seen in Fig.~\ref{ellipses}.

In the left frame of Fig.~\ref{lumfunc}, we plot as solid lines the best-fit luminosity functions as derived in Sec.~\ref{modeling} of this study (black), from the field population of MSPs, including uncertain distance determinations~\cite{Cholis:2014noa} (red), and using the MSP population in the globular cluster Tuc 47~\cite{Cholis:2014noa}, assuming a log-normal distribution in each case. Between luminosities of $10^{32}$ and $10^{34}$ erg/s, these three determinations are in fairly good agreement.  They depart, however, in their predictions for the number of very luminous MSPs.  In the right frame of Fig.~\ref{lumfunc}, we plot the best-fit values of $L_0$ and $\sigma_L$ for each of these three luminosity functions, as well as the 1 and 2$\sigma$ contours around the determination presented in Sec.~\ref{modeling} of this study, and the determination using field MSPs from Ref.~\cite{Cholis:2014noa}. When considering the determination from the fit in Sec.~\ref{modeling} of this study, recall from the discussion in Sec.~\ref{periods} that to reconcile this best-fit luminosity function with the observed variations in MSP periods, magnetic fields, and gamma-ray efficiencies, one must adopt $\sigma_L \gsim 1.5$, which is consistent with the range favored by the determinations presented in Ref.~\cite{Cholis:2014noa}.

The high luminosity end of the MSP gamma-ray luminosity function can be further constrained by considering the source PSR J1823-3021A.  This pulsar exhibits a period of 5.44 ms and a spin-down rate of $\dot{E}=8.28 \times 10^{35}$ erg/s. Importantly, PSR J1823-3021A resides in the globular cluster NGC 6624, making it possible to obtain a reliable distance measurement, $d=7.6 \pm0.4$ kpc~\cite{TheFermi-LAT:2013ssa}. This object is also highly luminous, $L_{\gamma}=7.0 \pm 1.0 \pm 0.8\times 10^{34}$~\cite{TheFermi-LAT:2013ssa}, outshining the total gamma-ray emission from most globular clusters (which are typically thought to contain large numbers of MSPs). 

Thus far in this study, we have focused on MSPs that are associated with the Galactic Plane, and have not taken into account PSR J1823-3021A, or any other sources residing within globular clusters. By comparing the gamma-ray luminosity of PSR J1823-3021A to the total luminosity from the 16 globular clusters observed by Fermi~\cite{Cholis:2014noa}, we find that more than 10\% of the total gamma-ray luminosity from these clusters comes from this single source. As this collection of 16 globular clusters contains approximately $\sim$2000 MSPs (using any of the luminosity functions shown in left frame of Fig.~\ref{lumfunc}), it is reasonable to expect that they collectively represent a fair sample of MSPs. If we attempt to correct for the possible bias resulting from Fermi not being able to detect faint globular clusters, this would only increase our estimate for the fraction of the total luminosity from globular clusters that comes from PSR J1823-3021A. We are also being conservative in assuming that all of the gamma-ray luminosity from these globular clusters originates from MSPs (their stacked spectrum suggests that this may not be the case~\cite{Cholis:2014noa}).

\begin{figure}
\includegraphics[width=3.40in,angle=0]{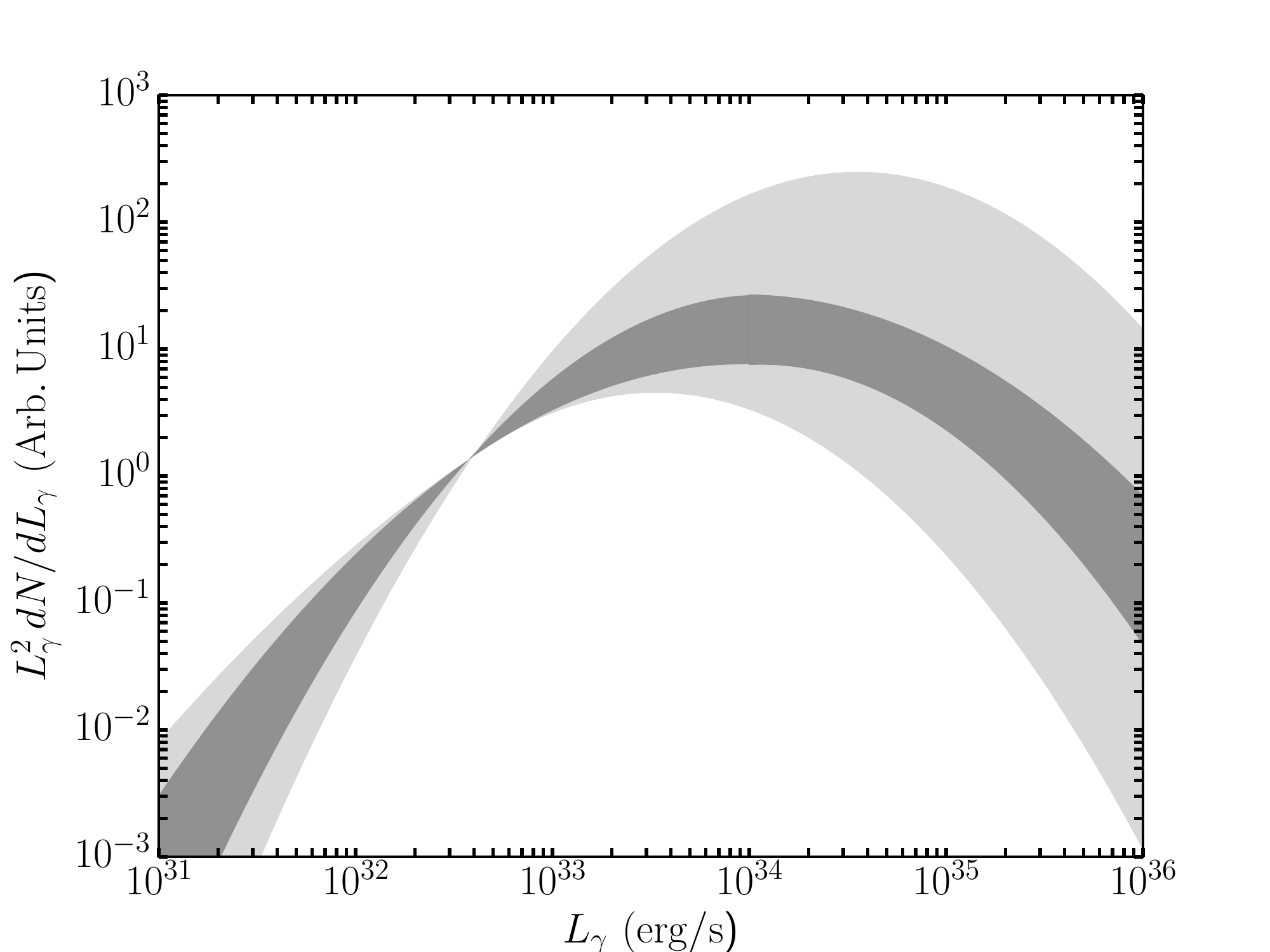} 
\caption{The range of MSP gamma-ray luminosity functions consistent at the 1 and 2$\sigma$ level to the combination of the population model fits described in Sec.~\ref{modeling} and the constraint based on PSR J1823-3021A described in Sec.~\ref{lumfuncsec}.}
\label{lumfuncfinal}
\end{figure}

If we adopt the best-fit luminosity function derived in Sec.~\ref{modeling} ($L_0=4.7 \times 10^{32}$ erg/s and $\sigma_L=1.4$), we predict that these 16 globular clusters should contain $\simeq$\,0.04 MSPs with $L_{\gamma} \ge 7 \times 10^{34}$ erg/s. Given that at least one such object is observed, this luminosity function model is excluded at the approximately 2$\sigma$ level. In the right frame of Fig.~\ref{lumfunc}, the grey shaded region represents the parameter space that is consistent (at the 95\% confidence level) with the observed luminosity of PSR J1823-3021A. In Fig.~\ref{lumfuncfinal}, we plot the 1 and 2$\sigma$ ranges for the luminosity function, as found when combining the likelihood function presented in Sec.~\ref{modeling} with the constraint based on PSR J1823-3021A.  Again, at luminosities between $10^{32}$ and $10^{34}$ erg/s, the range of allowed luminosity functions is fairly narrow, while a significant range of behaviors is found at higher luminosities, $L_{\gamma} \gsim 10^{34}$ erg/s.

\begin{table*}[t]
\begin{tabular}{|c|c|c|c|c|c|c|c|c|}
\hline
Model Name & $L_0$ (erg.s) &   $\sigma_L$  &  $N_{\rm MSP}$ & $N_{\rm MSP} (>10^{34})$ &  $N_{\rm MSP} (>10^{35})$ & $\langle L_{\gamma} \rangle$ (erg/s)& $N_{\rm det}$  \\
\hline \hline
A & 4.7$\times 10^{32}$       &        1.75 & 13,750     &   560.8   &  14.40 &          $2.16 \times 10^{33}$ & 19.9 \\
\hline
B  & 4.7$\times 10^{32}$       &        2.25 &  5187.2  &     456.2 &        44.36 &             $5.73 \times 10^{33}$ & 34.8  \\
\hline
C &   1.8$\times 10^{33}$       &        1.4  &   6212.0 &        689.7 &        12.49 &        $4.79\times 10^{33}$ & 21.4 \\
\hline
D &   5.5$\times 10^{33}$       &        1.4  &   2024.4 &       680.7 &     39.64 &          $1.47 \times 10^{34}$ & 36.8  \\
\hline \hline
Glob.~Clus.~\cite{Cholis:2014noa}  &   9.0$\times 10^{32}$       &        1.55  &  9982.4     &      602.8     &     11.02    &        $2.98\times 10^{33}$     &     18.8     \\
\hline
Field~\cite{Cholis:2014noa}  &   3.1$\times 10^{32}$       &        2.71  &  2681.5     &      272.8     &     46.25    &        $1.11\times 10^{34}$     &     31.7     \\
\hline \hline
\end{tabular}
\caption{A summary of the characteristics of a MSP population potentially responsible for the GeV excess, for six representative luminosity function models. For each model, we list the number of MSPs that are required to generate the observed intensity of the GeV excess, as well as the number of those MSPs with gamma-ray luminosities greater than $10^{34}$ and $10^{35}$ erg/s, and the mean luminosity in each model. The final column denotes the number of MSPs in the centrally located population that are predicted to be detectable by Fermi (and should appear within the 3FGL catalog).}
\label{MSP-table}
\end{table*}

\section{The Detectability of a MSP Population in the Inner Milky Way}
\label{innermw}

Motivated by the possibility that the GeV excess observed from the region surrounding the Galactic Center could potentially be generated by a population of unresolved MSPs~\cite{Hooper:2010mq,Hooper:2011ti,Abazajian:2012pn,Gordon:2013vta,Yuan:2014rca,Petrovic:2014xra,Brandt:2015ula,Bartels:2015aea}, we will utilize the luminosity function derived in this study to address this scenario.  To this end, we adopt a spherically symmetric spatial distribution of MSPs chosen to obtain the observed morphology of the excess: $dN/dV \propto r^{-2.4}$, out to $r_{\rm max}=3.1$ kpc. We then draw from a given luminosity function until the total luminosity from MSPs equals that of the GeV excess. 

In Table~\ref{MSP-table}, we provide a summary of our results, obtained for 6 representative choices of the gamma-ray luminosity function. In addition to the parameters describing each luminosity function ($L_0$, $\sigma_L$), we list the number of MSPs required to generate the observed intensity of the GeV excess, which varies between approximately 2000 and 14,000 over this range of models. Also listed is the number of those MSPs with gamma-ray luminosities greater than $10^{34}$ and $10^{35}$ erg/s, and the mean luminosity in each model.

\begin{figure}
\includegraphics[width=3.40in,angle=0]{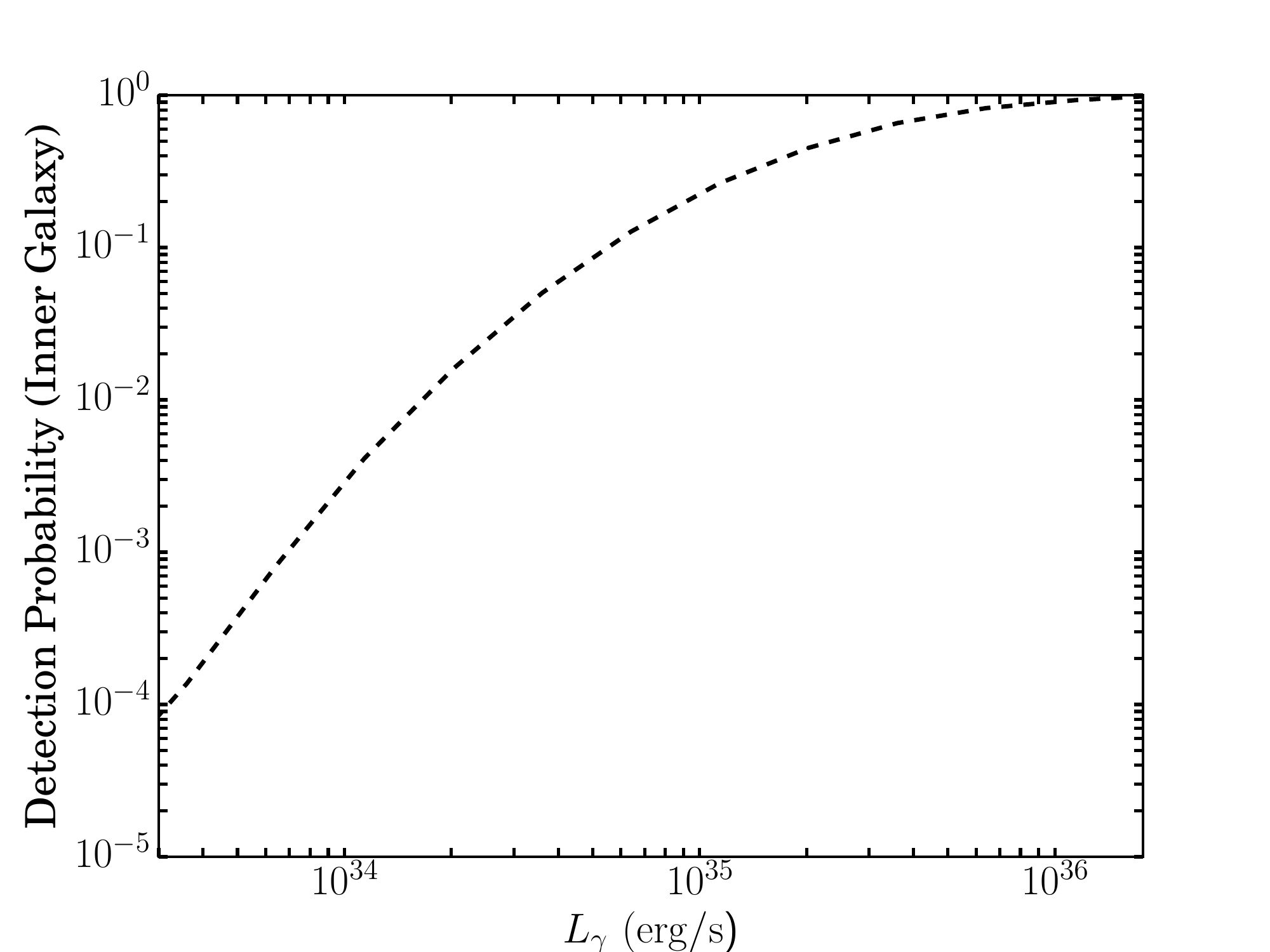} 
\caption{The probability that Fermi will have detected a MSP that is part of the population potentially responsible for the GeV excess, as a function of gamma-ray luminosity.}
\label{detectionprob}
\end{figure}

\begin{figure*}
\hspace{0.0cm}
\includegraphics[width=3.0in,angle=0]{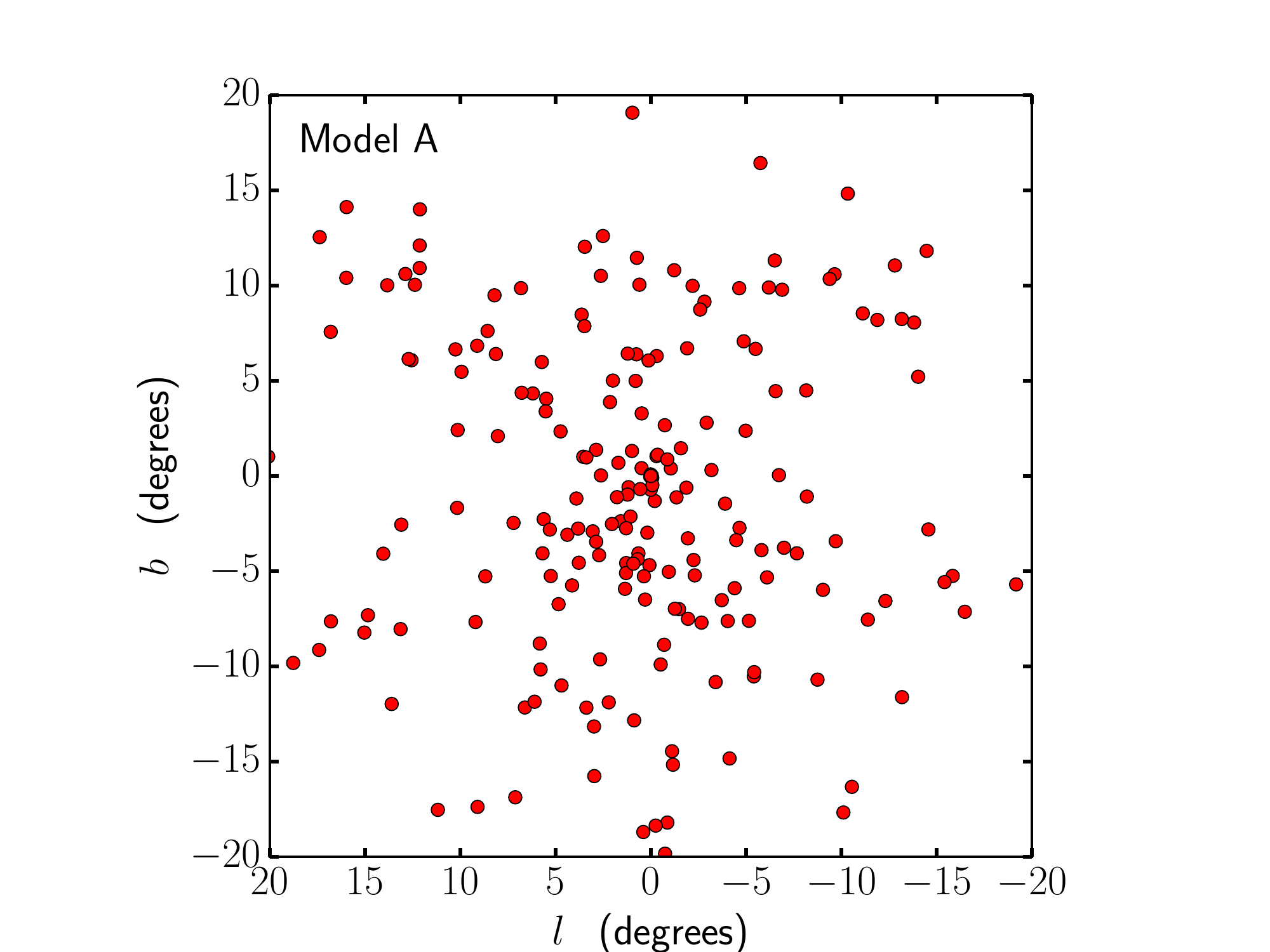} 
\hspace{-1.4cm}
\includegraphics[width=3.0in,angle=0]{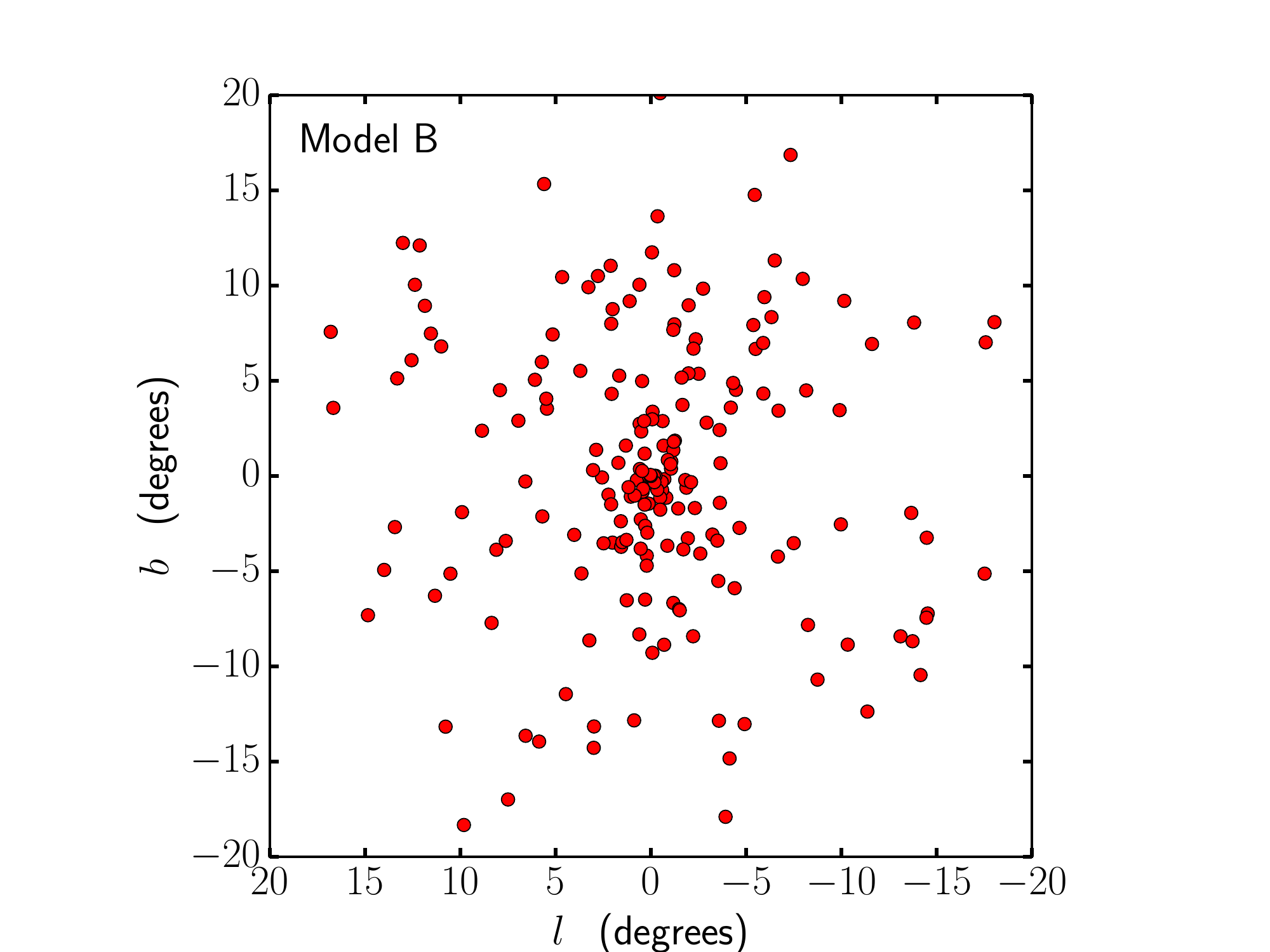}\\ 
\hspace{0.0cm}
\includegraphics[width=3.0in,angle=0]{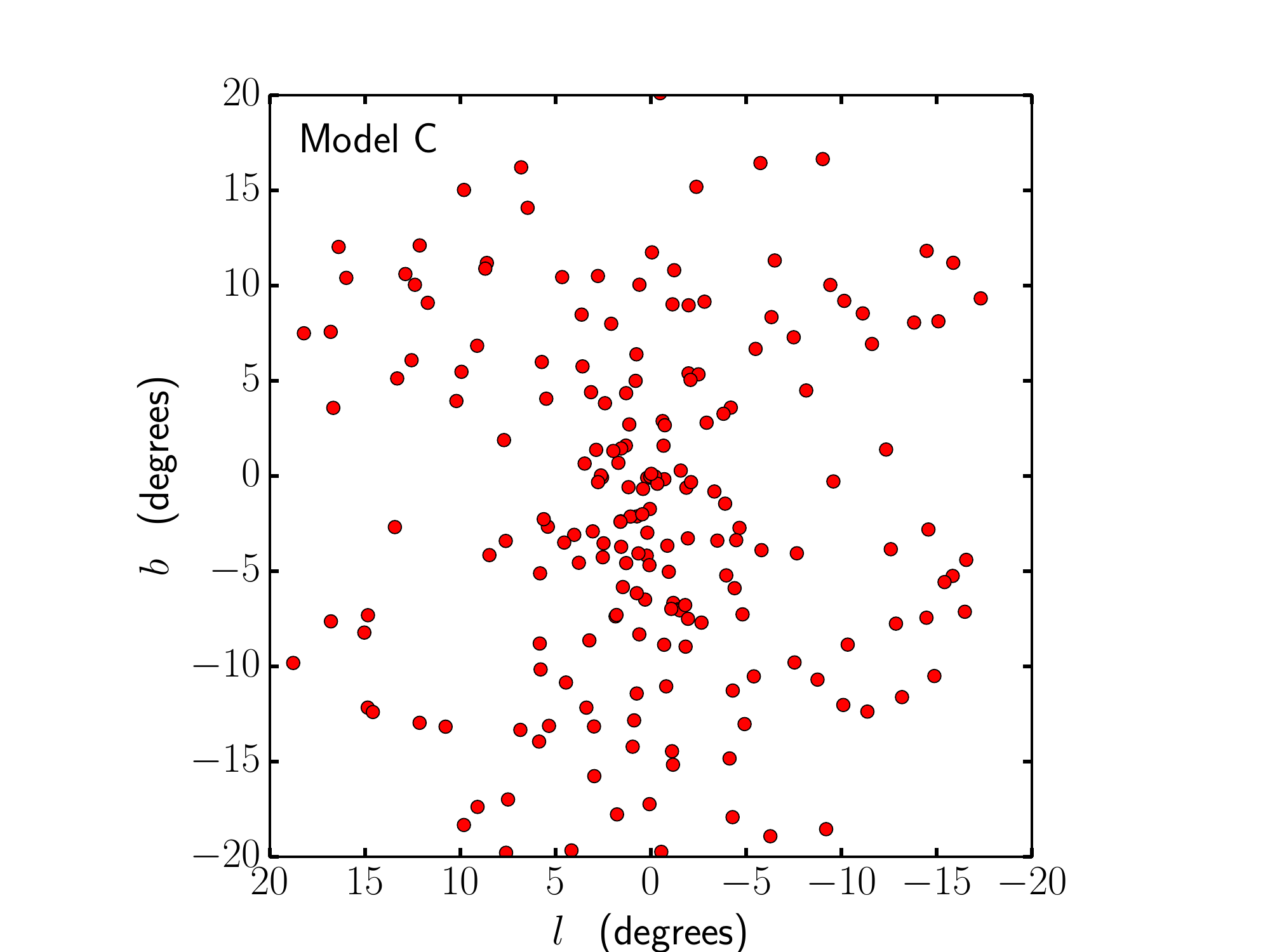} 
\hspace{-1.4cm}
\includegraphics[width=3.0in,angle=0]{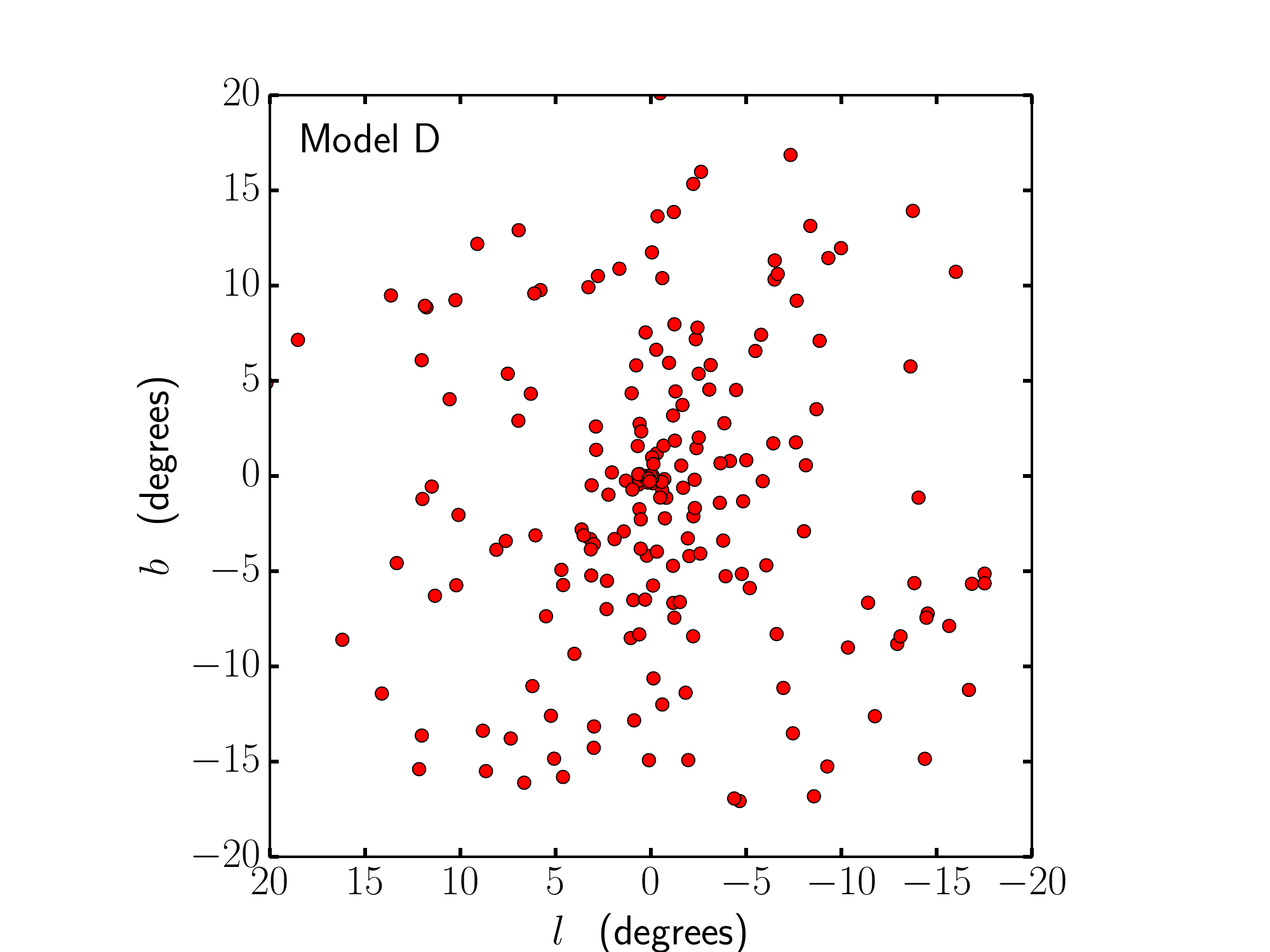} 
\caption{The angular distribution of a simulated sample of Inner Galaxy MSPs detectable by Fermi, for models A-D as described in Table~\ref{MSP-table}. In each frame, the number of detectable MSPs has been normalized to 200. Note that although the underlying MSP distribution is assumed to be spherically symmetric with respect to the Galactic Center, the higher backgrounds along the Galactic Plane suppress the number of sources that are detected within the region $b \lsim 2^{\circ}$.}
\label{regions}
\end{figure*}

To determine which members of the central MSP population would be detected by Fermi, we apply the same treatment for the detection probability as described in Sec.~\ref{modeling}, making use of the direction-dependent sensitivity map (Fig.~16, Ref.~\cite{TheFermi-LAT:2013ssa}), scaled to the best-fit normalization and variation found using the observed population of field MSPs. The probability of detecting a MSP that is part of the population potentially responsible for the GeV excess is shown in Fig.~\ref{detectionprob}, as a function of luminosity. This plot shows that most MSPs fainter than a few times $10^{34}$ erg/s will not be detected by Fermi (by which we mean they will not appear in the 3FGL catalog~\cite{Acero:2015hja}), and thus a very large fraction of this population will remain unresolved. Each of the models listed in Table~\ref{MSP-table} predict that Fermi should have detected between 18.8 and 36.8 MSPs if such a population is, in fact, responsible for generating the GeV excess. Alternatively, scanning over the range of luminosity functions shown in Fig.~\ref{lumfuncfinal}, we find that Fermi should have detected between approximately 15 and 43 sources associated with this MSP population. 

The authors of Ref.~\cite{Bartels:2015aea} have pointed out that there are 13 sources in the 3FGL catalog that lie within their region-of-interest ($2^{\circ}<|b|<12^{\circ}$ and $|l|<12^{\circ}$), are listed as unassociated with emission observed at other wavelengths, show no signs of variability on month-timescales, and have a spectrum that is roughly consistent with that expected from a MSP. It was subsequently shown in Ref.~\cite{Linden:2015qha}, however, that although the spectra of some of these sources (namely J1808.3-3357 and J1820.4-3217) do resemble that observed from other MSPs, the combined spectrum of these sources is not MSP-like, but instead resembles a power-law. This strongly suggests that a significant fraction of these sources are not, in fact, MSPs. Furthermore, 5 of these 13 sources fall within the region covered by the First Fermi-LAT Inner Galaxy point source catalog (1FIG), recently presented by the Fermi Collaboration~\cite{TheFermi-LAT:2015kwa}. Remarkably, 4 of the 5 of these MSP candidates are absent from this new catalog (3FGL J1740.5-2642, 3FGL J1740.8-1933, 3FGL J1759.2-3848, and 3FGL J1808.4-3519, see Table~3 of Ref.~\cite{TheFermi-LAT:2015kwa}). More generally, the Fermi Collaboration has suggested that a number of the 3FGL sources located near the Galactic Center, especially those located outside of the Galactic Plane, are likely to be associated with misattritubuted diffuse emission~\cite{TheFermi-LAT:2015kwa}. Furthermore, the distribution of sources in the 1FIG is significantly more concentrated around the Galactic Plane than was found in the 3FGL. The concentration of bright Inner Galaxy sources around the Galactic Plane was also identified in the analysis of Ref.~\cite{Lee:2015fea}.

In Fig.~\ref{regions}, we plot the simulated distribution of Inner Galaxy MSPs detectable by Fermi, for luminosity function models A, B, C, and D (see Table~\ref{MSP-table}). In each case, we have normalized the distribution to include 200 detectable MSPs. The most significant aspect of this plot is that comparatively few of the detectable MSPs are found near the Galactic Plane, due to the higher backgrounds and detection thresholds in that region of the sky. More quantitatively, the region along the Galactic Plane but excluding the Galactic Center ($|b|<2^{\circ}$, $|l|>2^{\circ}$) was found to contain only 30-40\% as many detectable MSPs as a similar region perpendicular to the Plane ($|l|<2^{\circ}$, $|b|>2^{\circ}$).


In light of this spectral and spatial information, it appears unlikely that more than half or so of the 13 sources listed in Ref.~\cite{Bartels:2015aea} could be part of a spherically symmetric MSP population.  In the same ROI, the range of luminosity function models shown in Fig.~\ref{lumfuncfinal} predict that Fermi should have detected between 9.2 and 25.8 MSPs. We thus conclude that if MSPs did account for the GeV excess, we should have expected Fermi to have detected significantly more MSP candidates than are contained in the 3FGL catalog.

While acknowledging the significant tension illustrated by this comparison, it is not our opinion that this firmly rules out the hypothesis that MSPs are responsible for the GeV excess. One way to evade this conclusion is to consider a population of MSPs in the Inner Galaxy that exhibits a different gamma-ray luminosity function than those observed in the field of the Galaxy, or in globular clusters.  For example, it was suggested in Ref.~\cite{Brandt:2015ula} that a sizable population of MSPs could have accumulated in the central stellar cluster and bulge of the Galaxy as the result of the tidal disruption of globular clusters. In such a scenario, the average MSP in the Inner Galaxy could be somewhat older than those observed elsewhere. 

The characteristic timescale for a pulsar to lose its rotational kinetic energy is given by:
\begin{eqnarray}
\tau \equiv \frac{E}{\dot{E}}= \frac{P}{2\dot{P}} \simeq 15.6 \, {\rm Gyr} \times \bigg(\frac{P}{5 \, {\rm ms}}\bigg)^2 \, \bigg(\frac{1.6\times 10^8 \, {\rm G}}{B}\bigg)^2.
\end{eqnarray}
Given that this is of the same order of magnitude as the age of many globular clusters, it seems plausible that a very old MSP population (without the young members that might have formed more recently in the Galactic Disk or within intact globular clusters) might exhibit slightly longer periods, and thus modestly lower gamma-ray luminosities, than those MSPs observed elsewhere.

\section{Summary and Conclusions}
\label{conclusions}
                   
The viability of millisecond pulsars (MSPs) as the source of the GeV Galactic Center excess depends critically on the fraction of these objects that are luminous enough to be resolved by Fermi as individual point sources. In this paper, we revisited the MSP gamma-ray luminosity function. Making use of the observed distribution of gamma-ray MSPs on the sky and in flux, we have constrained the characteristics of the Milky Way's MSP population, finding a spatial distribution and distribution of periods that is in good agreement with those found in previous studies. Combining these results with information pertaining to the MSP population observed in globular clusters, we have produced a robust determination of the luminosity function, which does not rely on any uncertain distance measurements. 

Using the luminosity function derived in this study, we predict that if the GeV excess originates from MSPs, Fermi should have already detected between 15 and 43 members of the Inner Galaxy population. Furthermore, due to the formidable backgrounds found along the Galactic Plane, the detectable MSPs are predicted to reside preferentially at $|b| \gsim 2^{\circ}$. The predicted quantity and distribution of detectable MSPs is inconsistent with the MSP candidates contained within the 3FGL and 1FIG catalogs.  If MSPs are responsible for the gamma-ray signal observed from the Inner Galaxy, they must be systematically less luminous than those pulsars observed in the Galactic Plane or within globular clusters.   
                   
\bigskip    
\bigskip
\vskip 0.6 in

\section*{Acknowledgments}  
We would like to thank Tim Linden, Christoph Weniger, Richard Bartels, Alex Drlica-Wagner, and Timothy Brandt for valuable discussions. GM is supported by the Fermilab Graduate Student Research Program in Theoretical Physics and in part by the National Research Foundation of South Africa, Grant No.~88614. This work has been supported by the US Department of Energy. 

\vskip 0.05in

\bibliography{msp2015}

\begin{thebibliography}{72}
\expandafter\ifx\csname natexlab\endcsname\relax\def\natexlab#1{#1}\fi
\expandafter\ifx\csname bibnamefont\endcsname\relax
  \def\bibnamefont#1{#1}\fi
\expandafter\ifx\csname bibfnamefont\endcsname\relax
  \def\bibfnamefont#1{#1}\fi
\expandafter\ifx\csname citenamefont\endcsname\relax
  \def\citenamefont#1{#1}\fi
\expandafter\ifx\csname url\endcsname\relax
  \def\url#1{\texttt{#1}}\fi
\expandafter\ifx\csname urlprefix\endcsname\relax\def\urlprefix{URL }\fi
\providecommand{\bibinfo}[2]{#2}
\providecommand{\eprint}[2][]{\url{#2}}

\bibitem[{\citenamefont{{Abdo} et~al.}(2009)}]{2009Sci...325..848A}
\bibinfo{author}{\bibfnamefont{A.~A.} \bibnamefont{{Abdo}}}
  \bibnamefont{et~al.}, \bibinfo{journal}{Science}
  \textbf{\bibinfo{volume}{325}}, \bibinfo{pages}{848} (\bibinfo{year}{2009}).

\bibitem[{\citenamefont{Abdo et~al.}(2013)}]{TheFermi-LAT:2013ssa}
\bibinfo{author}{\bibfnamefont{A.}~\bibnamefont{Abdo}} \bibnamefont{et~al.}
  (\bibinfo{collaboration}{The Fermi-LAT collaboration}),
  \bibinfo{journal}{Astrophys.J.Suppl.} \textbf{\bibinfo{volume}{208}},
  \bibinfo{pages}{17} (\bibinfo{year}{2013}), \eprint{1305.4385}.

\bibitem[{\citenamefont{{Alpar} et~al.}(1982)\citenamefont{{Alpar}, {Cheng},
  {Ruderman}, and {Shaham}}}]{1982Natur.300..728A}
\bibinfo{author}{\bibfnamefont{M.~A.} \bibnamefont{{Alpar}}},
  \bibinfo{author}{\bibfnamefont{A.~F.} \bibnamefont{{Cheng}}},
  \bibinfo{author}{\bibfnamefont{M.~A.} \bibnamefont{{Ruderman}}},
  \bibnamefont{and} \bibinfo{author}{\bibfnamefont{J.}~\bibnamefont{{Shaham}}},
  \bibinfo{journal}{\nat} \textbf{\bibinfo{volume}{300}}, \bibinfo{pages}{728}
  (\bibinfo{year}{1982}).

\bibitem[{\citenamefont{{Phinney} and {Kulkarni}}(1994)}]{1994ARA&A..32..591P}
\bibinfo{author}{\bibfnamefont{E.~S.} \bibnamefont{{Phinney}}}
  \bibnamefont{and} \bibinfo{author}{\bibfnamefont{S.~R.}
  \bibnamefont{{Kulkarni}}}, \bibinfo{journal}{\araa}
  \textbf{\bibinfo{volume}{32}}, \bibinfo{pages}{591} (\bibinfo{year}{1994}).

\bibitem[{\citenamefont{Lorimer}(2001)}]{Lorimer:2001vd}
\bibinfo{author}{\bibfnamefont{D.~R.} \bibnamefont{Lorimer}},
  \bibinfo{journal}{Living Rev. Rel.} \textbf{\bibinfo{volume}{4}},
  \bibinfo{pages}{5} (\bibinfo{year}{2001}), \eprint{astro-ph/0104388}.

\bibitem[{\citenamefont{Lorimer}(2008)}]{Lorimer:2008se}
\bibinfo{author}{\bibfnamefont{D.~R.} \bibnamefont{Lorimer}},
  \bibinfo{journal}{Living Rev. Rel.} \textbf{\bibinfo{volume}{11}},
  \bibinfo{pages}{8} (\bibinfo{year}{2008}), \eprint{0811.0762}.

\bibitem[{\citenamefont{{Kiziltan} and {Thorsett}}(2010)}]{2010ApJ...715..335K}
\bibinfo{author}{\bibfnamefont{B.}~\bibnamefont{{Kiziltan}}} \bibnamefont{and}
  \bibinfo{author}{\bibfnamefont{S.~E.} \bibnamefont{{Thorsett}}},
  \bibinfo{journal}{\apj} \textbf{\bibinfo{volume}{715}}, \bibinfo{pages}{335}
  (\bibinfo{year}{2010}), \eprint{0909.1562}.

\bibitem[{\citenamefont{Guillemot et~al.}(2012)\citenamefont{Guillemot,
  Johnson, Venter, Kerr, Pancrazi et~al.}}]{Guillemot:2011th}
\bibinfo{author}{\bibfnamefont{L.}~\bibnamefont{Guillemot}},
  \bibinfo{author}{\bibfnamefont{T.}~\bibnamefont{Johnson}},
  \bibinfo{author}{\bibfnamefont{C.}~\bibnamefont{Venter}},
  \bibinfo{author}{\bibfnamefont{M.}~\bibnamefont{Kerr}},
  \bibinfo{author}{\bibfnamefont{B.}~\bibnamefont{Pancrazi}},
  \bibnamefont{et~al.}, \bibinfo{journal}{Astrophys.J.}
  \textbf{\bibinfo{volume}{744}}, \bibinfo{pages}{33} (\bibinfo{year}{2012}),
  \eprint{1110.1271}.

\bibitem[{\citenamefont{Faucher-Giguere and
  Loeb}(2010)}]{FaucherGiguere:2009df}
\bibinfo{author}{\bibfnamefont{C.~A.} \bibnamefont{Faucher-Giguere}}
  \bibnamefont{and} \bibinfo{author}{\bibfnamefont{A.}~\bibnamefont{Loeb}},
  \bibinfo{journal}{JCAP} \textbf{\bibinfo{volume}{1001}}, \bibinfo{pages}{005}
  (\bibinfo{year}{2010}), \eprint{0904.3102}.

\bibitem[{\citenamefont{{Sturner} and {Dermer}}(1996)}]{1996ApJ...461..872S}
\bibinfo{author}{\bibfnamefont{S.~J.} \bibnamefont{{Sturner}}}
  \bibnamefont{and} \bibinfo{author}{\bibfnamefont{C.~D.}
  \bibnamefont{{Dermer}}}, \bibinfo{journal}{\apj}
  \textbf{\bibinfo{volume}{461}}, \bibinfo{pages}{872} (\bibinfo{year}{1996}).

\bibitem[{\citenamefont{Hooper et~al.}(2013)\citenamefont{Hooper, Cholis,
  Linden, Siegal-Gaskins, and Slatyer}}]{Hooper:2013nhl}
\bibinfo{author}{\bibfnamefont{D.}~\bibnamefont{Hooper}},
  \bibinfo{author}{\bibfnamefont{I.}~\bibnamefont{Cholis}},
  \bibinfo{author}{\bibfnamefont{T.}~\bibnamefont{Linden}},
  \bibinfo{author}{\bibfnamefont{J.}~\bibnamefont{Siegal-Gaskins}},
  \bibnamefont{and} \bibinfo{author}{\bibfnamefont{T.~R.}
  \bibnamefont{Slatyer}}, \bibinfo{journal}{Phys.Rev.}
  \textbf{\bibinfo{volume}{D88}}, \bibinfo{pages}{083009}
  (\bibinfo{year}{2013}), \eprint{1305.0830}.

\bibitem[{\citenamefont{Grégoire and Knödlseder}(2013)}]{Gregoire:2013yta}
\bibinfo{author}{\bibfnamefont{T.}~\bibnamefont{Grégoire}} \bibnamefont{and}
  \bibinfo{author}{\bibfnamefont{J.}~\bibnamefont{Knödlseder}},
  \bibinfo{journal}{Astron. Astrophys.} \textbf{\bibinfo{volume}{554}},
  \bibinfo{pages}{A62} (\bibinfo{year}{2013}), \eprint{1305.1584}.

\bibitem[{\citenamefont{Calore et~al.}(2014)\citenamefont{Calore, Di~Mauro, and
  Donato}}]{Calore:2014oga}
\bibinfo{author}{\bibfnamefont{F.}~\bibnamefont{Calore}},
  \bibinfo{author}{\bibfnamefont{M.}~\bibnamefont{Di~Mauro}}, \bibnamefont{and}
  \bibinfo{author}{\bibfnamefont{F.}~\bibnamefont{Donato}}
  (\bibinfo{year}{2014}), \eprint{1406.2706}.

\bibitem[{\citenamefont{Siegal-Gaskins
  et~al.}(2010)\citenamefont{Siegal-Gaskins, Reesman, Pavlidou, Profumo, and
  Walker}}]{SiegalGaskins:2010mp}
\bibinfo{author}{\bibfnamefont{J.~M.} \bibnamefont{Siegal-Gaskins}},
  \bibinfo{author}{\bibfnamefont{R.}~\bibnamefont{Reesman}},
  \bibinfo{author}{\bibfnamefont{V.}~\bibnamefont{Pavlidou}},
  \bibinfo{author}{\bibfnamefont{S.}~\bibnamefont{Profumo}}, \bibnamefont{and}
  \bibinfo{author}{\bibfnamefont{T.~P.} \bibnamefont{Walker}}
  (\bibinfo{year}{2010}), \eprint{1011.5501}.

\bibitem[{\citenamefont{Goodenough and Hooper}(2009)}]{Goodenough:2009gk}
\bibinfo{author}{\bibfnamefont{L.}~\bibnamefont{Goodenough}} \bibnamefont{and}
  \bibinfo{author}{\bibfnamefont{D.}~\bibnamefont{Hooper}}
  (\bibinfo{year}{2009}), \eprint{0910.2998}.

\bibitem[{\citenamefont{Hooper and Goodenough}(2011)}]{Hooper:2010mq}
\bibinfo{author}{\bibfnamefont{D.}~\bibnamefont{Hooper}} \bibnamefont{and}
  \bibinfo{author}{\bibfnamefont{L.}~\bibnamefont{Goodenough}},
  \bibinfo{journal}{Phys.Lett.} \textbf{\bibinfo{volume}{B697}},
  \bibinfo{pages}{412} (\bibinfo{year}{2011}), \eprint{1010.2752}.

\bibitem[{\citenamefont{Hooper and Linden}(2011)}]{Hooper:2011ti}
\bibinfo{author}{\bibfnamefont{D.}~\bibnamefont{Hooper}} \bibnamefont{and}
  \bibinfo{author}{\bibfnamefont{T.}~\bibnamefont{Linden}},
  \bibinfo{journal}{Phys.Rev.} \textbf{\bibinfo{volume}{D84}},
  \bibinfo{pages}{123005} (\bibinfo{year}{2011}), \eprint{1110.0006}.

\bibitem[{\citenamefont{Abazajian and Kaplinghat}(2012)}]{Abazajian:2012pn}
\bibinfo{author}{\bibfnamefont{K.~N.} \bibnamefont{Abazajian}}
  \bibnamefont{and}
  \bibinfo{author}{\bibfnamefont{M.}~\bibnamefont{Kaplinghat}},
  \bibinfo{journal}{Phys.Rev.} \textbf{\bibinfo{volume}{D86}},
  \bibinfo{pages}{083511} (\bibinfo{year}{2012}), \eprint{1207.6047}.

\bibitem[{\citenamefont{Gordon and Macias}(2013)}]{Gordon:2013vta}
\bibinfo{author}{\bibfnamefont{C.}~\bibnamefont{Gordon}} \bibnamefont{and}
  \bibinfo{author}{\bibfnamefont{O.}~\bibnamefont{Macias}},
  \bibinfo{journal}{Phys.Rev.} \textbf{\bibinfo{volume}{D88}},
  \bibinfo{pages}{083521} (\bibinfo{year}{2013}), \eprint{1306.5725}.

\bibitem[{\citenamefont{Hooper and Slatyer}(2013)}]{Hooper:2013rwa}
\bibinfo{author}{\bibfnamefont{D.}~\bibnamefont{Hooper}} \bibnamefont{and}
  \bibinfo{author}{\bibfnamefont{T.~R.} \bibnamefont{Slatyer}},
  \bibinfo{journal}{Phys.Dark Univ.} \textbf{\bibinfo{volume}{2}},
  \bibinfo{pages}{118} (\bibinfo{year}{2013}), \eprint{1302.6589}.

\bibitem[{\citenamefont{Daylan et~al.}(2014)\citenamefont{Daylan, Finkbeiner,
  Hooper, Linden, Portillo et~al.}}]{Daylan:2014rsa}
\bibinfo{author}{\bibfnamefont{T.}~\bibnamefont{Daylan}},
  \bibinfo{author}{\bibfnamefont{D.~P.} \bibnamefont{Finkbeiner}},
  \bibinfo{author}{\bibfnamefont{D.}~\bibnamefont{Hooper}},
  \bibinfo{author}{\bibfnamefont{T.}~\bibnamefont{Linden}},
  \bibinfo{author}{\bibfnamefont{S.~K.~N.} \bibnamefont{Portillo}},
  \bibnamefont{et~al.} (\bibinfo{year}{2014}), \eprint{1402.6703}.

\bibitem[{\citenamefont{Calore et~al.}(2015)\citenamefont{Calore, Cholis, and
  Weniger}}]{Calore:2014xka}
\bibinfo{author}{\bibfnamefont{F.}~\bibnamefont{Calore}},
  \bibinfo{author}{\bibfnamefont{I.}~\bibnamefont{Cholis}}, \bibnamefont{and}
  \bibinfo{author}{\bibfnamefont{C.}~\bibnamefont{Weniger}},
  \bibinfo{journal}{JCAP} \textbf{\bibinfo{volume}{1503}}, \bibinfo{pages}{038}
  (\bibinfo{year}{2015}), \eprint{1409.0042}.

\bibitem[{\citenamefont{Ajello et~al.}(2015)}]{TheFermi-LAT:2015kwa}
\bibinfo{author}{\bibfnamefont{M.}~\bibnamefont{Ajello}} \bibnamefont{et~al.}
  (\bibinfo{collaboration}{Fermi-LAT}) (\bibinfo{year}{2015}),
  \eprint{1511.02938}.

\bibitem[{\citenamefont{Abdullah et~al.}(2014)\citenamefont{Abdullah, DiFranzo,
  Rajaraman, Tait, Tanedo, and Wijangco}}]{Abdullah:2014lla}
\bibinfo{author}{\bibfnamefont{M.}~\bibnamefont{Abdullah}},
  \bibinfo{author}{\bibfnamefont{A.}~\bibnamefont{DiFranzo}},
  \bibinfo{author}{\bibfnamefont{A.}~\bibnamefont{Rajaraman}},
  \bibinfo{author}{\bibfnamefont{T.~M.~P.} \bibnamefont{Tait}},
  \bibinfo{author}{\bibfnamefont{P.}~\bibnamefont{Tanedo}}, \bibnamefont{and}
  \bibinfo{author}{\bibfnamefont{A.~M.} \bibnamefont{Wijangco}},
  \bibinfo{journal}{Phys. Rev.} \textbf{\bibinfo{volume}{D90}},
  \bibinfo{pages}{035004} (\bibinfo{year}{2014}), \eprint{1404.6528}.

\bibitem[{\citenamefont{Ipek et~al.}(2014)\citenamefont{Ipek, McKeen, and
  Nelson}}]{Ipek:2014gua}
\bibinfo{author}{\bibfnamefont{S.}~\bibnamefont{Ipek}},
  \bibinfo{author}{\bibfnamefont{D.}~\bibnamefont{McKeen}}, \bibnamefont{and}
  \bibinfo{author}{\bibfnamefont{A.~E.} \bibnamefont{Nelson}},
  \bibinfo{journal}{Phys. Rev.} \textbf{\bibinfo{volume}{D90}},
  \bibinfo{pages}{055021} (\bibinfo{year}{2014}), \eprint{1404.3716}.

\bibitem[{\citenamefont{Izaguirre et~al.}(2014)\citenamefont{Izaguirre,
  Krnjaic, and Shuve}}]{Izaguirre:2014vva}
\bibinfo{author}{\bibfnamefont{E.}~\bibnamefont{Izaguirre}},
  \bibinfo{author}{\bibfnamefont{G.}~\bibnamefont{Krnjaic}}, \bibnamefont{and}
  \bibinfo{author}{\bibfnamefont{B.}~\bibnamefont{Shuve}},
  \bibinfo{journal}{Phys. Rev.} \textbf{\bibinfo{volume}{D90}},
  \bibinfo{pages}{055002} (\bibinfo{year}{2014}), \eprint{1404.2018}.

\bibitem[{\citenamefont{Agrawal et~al.}(2014)\citenamefont{Agrawal, Batell,
  Hooper, and Lin}}]{Agrawal:2014una}
\bibinfo{author}{\bibfnamefont{P.}~\bibnamefont{Agrawal}},
  \bibinfo{author}{\bibfnamefont{B.}~\bibnamefont{Batell}},
  \bibinfo{author}{\bibfnamefont{D.}~\bibnamefont{Hooper}}, \bibnamefont{and}
  \bibinfo{author}{\bibfnamefont{T.}~\bibnamefont{Lin}},
  \bibinfo{journal}{Phys. Rev.} \textbf{\bibinfo{volume}{D90}},
  \bibinfo{pages}{063512} (\bibinfo{year}{2014}), \eprint{1404.1373}.

\bibitem[{\citenamefont{Berlin et~al.}(2014)\citenamefont{Berlin, Hooper, and
  McDermott}}]{Berlin:2014tja}
\bibinfo{author}{\bibfnamefont{A.}~\bibnamefont{Berlin}},
  \bibinfo{author}{\bibfnamefont{D.}~\bibnamefont{Hooper}}, \bibnamefont{and}
  \bibinfo{author}{\bibfnamefont{S.~D.} \bibnamefont{McDermott}},
  \bibinfo{journal}{Phys. Rev.} \textbf{\bibinfo{volume}{D89}},
  \bibinfo{pages}{115022} (\bibinfo{year}{2014}), \eprint{1404.0022}.

\bibitem[{\citenamefont{Alves et~al.}(2014)\citenamefont{Alves, Profumo,
  Queiroz, and Shepherd}}]{Alves:2014yha}
\bibinfo{author}{\bibfnamefont{A.}~\bibnamefont{Alves}},
  \bibinfo{author}{\bibfnamefont{S.}~\bibnamefont{Profumo}},
  \bibinfo{author}{\bibfnamefont{F.~S.} \bibnamefont{Queiroz}},
  \bibnamefont{and} \bibinfo{author}{\bibfnamefont{W.}~\bibnamefont{Shepherd}},
  \bibinfo{journal}{Phys. Rev.} \textbf{\bibinfo{volume}{D90}},
  \bibinfo{pages}{115003} (\bibinfo{year}{2014}), \eprint{1403.5027}.

\bibitem[{\citenamefont{Boehm et~al.}(2014)\citenamefont{Boehm, Dolan, McCabe,
  Spannowsky, and Wallace}}]{Boehm:2014hva}
\bibinfo{author}{\bibfnamefont{C.}~\bibnamefont{Boehm}},
  \bibinfo{author}{\bibfnamefont{M.~J.} \bibnamefont{Dolan}},
  \bibinfo{author}{\bibfnamefont{C.}~\bibnamefont{McCabe}},
  \bibinfo{author}{\bibfnamefont{M.}~\bibnamefont{Spannowsky}},
  \bibnamefont{and} \bibinfo{author}{\bibfnamefont{C.~J.}
  \bibnamefont{Wallace}}, \bibinfo{journal}{JCAP}
  \textbf{\bibinfo{volume}{1405}}, \bibinfo{pages}{009} (\bibinfo{year}{2014}),
  \eprint{1401.6458}.

\bibitem[{\citenamefont{Martin et~al.}(2014)\citenamefont{Martin, Shelton, and
  Unwin}}]{Martin:2014sxa}
\bibinfo{author}{\bibfnamefont{A.}~\bibnamefont{Martin}},
  \bibinfo{author}{\bibfnamefont{J.}~\bibnamefont{Shelton}}, \bibnamefont{and}
  \bibinfo{author}{\bibfnamefont{J.}~\bibnamefont{Unwin}},
  \bibinfo{journal}{Phys. Rev.} \textbf{\bibinfo{volume}{D90}},
  \bibinfo{pages}{103513} (\bibinfo{year}{2014}), \eprint{1405.0272}.

\bibitem[{\citenamefont{Huang et~al.}(2014)\citenamefont{Huang, Liu, Wang, and
  Yu}}]{Huang:2014cla}
\bibinfo{author}{\bibfnamefont{J.}~\bibnamefont{Huang}},
  \bibinfo{author}{\bibfnamefont{T.}~\bibnamefont{Liu}},
  \bibinfo{author}{\bibfnamefont{L.-T.} \bibnamefont{Wang}}, \bibnamefont{and}
  \bibinfo{author}{\bibfnamefont{F.}~\bibnamefont{Yu}}, \bibinfo{journal}{Phys.
  Rev.} \textbf{\bibinfo{volume}{D90}}, \bibinfo{pages}{115006}
  (\bibinfo{year}{2014}), \eprint{1407.0038}.

\bibitem[{\citenamefont{Cerdeño et~al.}(2014)\citenamefont{Cerdeño, Peiró,
  and Robles}}]{Cerdeno:2014cda}
\bibinfo{author}{\bibfnamefont{D.~G.} \bibnamefont{Cerdeño}},
  \bibinfo{author}{\bibfnamefont{M.}~\bibnamefont{Peiró}}, \bibnamefont{and}
  \bibinfo{author}{\bibfnamefont{S.}~\bibnamefont{Robles}},
  \bibinfo{journal}{JCAP} \textbf{\bibinfo{volume}{1408}}, \bibinfo{pages}{005}
  (\bibinfo{year}{2014}), \eprint{1404.2572}.

\bibitem[{\citenamefont{Okada and Seto}(2014)}]{Okada:2013bna}
\bibinfo{author}{\bibfnamefont{N.}~\bibnamefont{Okada}} \bibnamefont{and}
  \bibinfo{author}{\bibfnamefont{O.}~\bibnamefont{Seto}},
  \bibinfo{journal}{Phys. Rev.} \textbf{\bibinfo{volume}{D89}},
  \bibinfo{pages}{043525} (\bibinfo{year}{2014}), \eprint{1310.5991}.

\bibitem[{\citenamefont{Freese et~al.}(2015)\citenamefont{Freese, Lopez, Shah,
  and Shakya}}]{Freese:2015ysa}
\bibinfo{author}{\bibfnamefont{K.}~\bibnamefont{Freese}},
  \bibinfo{author}{\bibfnamefont{A.}~\bibnamefont{Lopez}},
  \bibinfo{author}{\bibfnamefont{N.~R.} \bibnamefont{Shah}}, \bibnamefont{and}
  \bibinfo{author}{\bibfnamefont{B.}~\bibnamefont{Shakya}}
  (\bibinfo{year}{2015}), \eprint{1509.05076}.

\bibitem[{\citenamefont{Fonseca et~al.}(2015)\citenamefont{Fonseca, Necib, and
  Thaler}}]{Fonseca:2015rwa}
\bibinfo{author}{\bibfnamefont{N.}~\bibnamefont{Fonseca}},
  \bibinfo{author}{\bibfnamefont{L.}~\bibnamefont{Necib}}, \bibnamefont{and}
  \bibinfo{author}{\bibfnamefont{J.}~\bibnamefont{Thaler}}
  (\bibinfo{year}{2015}), \eprint{1507.08295}.

\bibitem[{\citenamefont{Bertone et~al.}(2015)\citenamefont{Bertone, Calore,
  Caron, de~Austri, Kim, Trotta, and Weniger}}]{Bertone:2015tza}
\bibinfo{author}{\bibfnamefont{G.}~\bibnamefont{Bertone}},
  \bibinfo{author}{\bibfnamefont{F.}~\bibnamefont{Calore}},
  \bibinfo{author}{\bibfnamefont{S.}~\bibnamefont{Caron}},
  \bibinfo{author}{\bibfnamefont{R.~R.} \bibnamefont{de~Austri}},
  \bibinfo{author}{\bibfnamefont{J.~S.} \bibnamefont{Kim}},
  \bibinfo{author}{\bibfnamefont{R.}~\bibnamefont{Trotta}}, \bibnamefont{and}
  \bibinfo{author}{\bibfnamefont{C.}~\bibnamefont{Weniger}}
  (\bibinfo{year}{2015}), \eprint{1507.07008}.

\bibitem[{\citenamefont{Cline et~al.}(2015)\citenamefont{Cline, Dupuis, Liu,
  and Xue}}]{Cline:2015qha}
\bibinfo{author}{\bibfnamefont{J.~M.} \bibnamefont{Cline}},
  \bibinfo{author}{\bibfnamefont{G.}~\bibnamefont{Dupuis}},
  \bibinfo{author}{\bibfnamefont{Z.}~\bibnamefont{Liu}}, \bibnamefont{and}
  \bibinfo{author}{\bibfnamefont{W.}~\bibnamefont{Xue}},
  \bibinfo{journal}{Phys. Rev.} \textbf{\bibinfo{volume}{D91}},
  \bibinfo{pages}{115010} (\bibinfo{year}{2015}), \eprint{1503.08213}.

\bibitem[{\citenamefont{Berlin et~al.}(2015)\citenamefont{Berlin, Gori, Lin,
  and Wang}}]{Berlin:2015wwa}
\bibinfo{author}{\bibfnamefont{A.}~\bibnamefont{Berlin}},
  \bibinfo{author}{\bibfnamefont{S.}~\bibnamefont{Gori}},
  \bibinfo{author}{\bibfnamefont{T.}~\bibnamefont{Lin}}, \bibnamefont{and}
  \bibinfo{author}{\bibfnamefont{L.-T.} \bibnamefont{Wang}},
  \bibinfo{journal}{Phys. Rev.} \textbf{\bibinfo{volume}{D92}},
  \bibinfo{pages}{015005} (\bibinfo{year}{2015}), \eprint{1502.06000}.

\bibitem[{\citenamefont{Achterberg et~al.}(2015)\citenamefont{Achterberg,
  Amoroso, Caron, Hendriks, Ruiz~de Austri, and Weniger}}]{Caron:2015wda}
\bibinfo{author}{\bibfnamefont{A.}~\bibnamefont{Achterberg}},
  \bibinfo{author}{\bibfnamefont{S.}~\bibnamefont{Amoroso}},
  \bibinfo{author}{\bibfnamefont{S.}~\bibnamefont{Caron}},
  \bibinfo{author}{\bibfnamefont{L.}~\bibnamefont{Hendriks}},
  \bibinfo{author}{\bibfnamefont{R.}~\bibnamefont{Ruiz~de Austri}},
  \bibnamefont{and} \bibinfo{author}{\bibfnamefont{C.}~\bibnamefont{Weniger}},
  \bibinfo{journal}{JCAP} \textbf{\bibinfo{volume}{1508}}, \bibinfo{pages}{006}
  (\bibinfo{year}{2015}), \eprint{1502.05703}.

\bibitem[{\citenamefont{Cerdeno et~al.}(2015)\citenamefont{Cerdeno, Peiro, and
  Robles}}]{Cerdeno:2015ega}
\bibinfo{author}{\bibfnamefont{D.~G.} \bibnamefont{Cerdeno}},
  \bibinfo{author}{\bibfnamefont{M.}~\bibnamefont{Peiro}}, \bibnamefont{and}
  \bibinfo{author}{\bibfnamefont{S.}~\bibnamefont{Robles}},
  \bibinfo{journal}{Phys. Rev.} \textbf{\bibinfo{volume}{D91}},
  \bibinfo{pages}{123530} (\bibinfo{year}{2015}), \eprint{1501.01296}.

\bibitem[{\citenamefont{Liu et~al.}(2015)\citenamefont{Liu, Weiner, and
  Xue}}]{Liu:2014cma}
\bibinfo{author}{\bibfnamefont{J.}~\bibnamefont{Liu}},
  \bibinfo{author}{\bibfnamefont{N.}~\bibnamefont{Weiner}}, \bibnamefont{and}
  \bibinfo{author}{\bibfnamefont{W.}~\bibnamefont{Xue}},
  \bibinfo{journal}{JHEP} \textbf{\bibinfo{volume}{08}}, \bibinfo{pages}{050}
  (\bibinfo{year}{2015}), \eprint{1412.1485}.

\bibitem[{\citenamefont{Hooper}(2015)}]{Hooper:2014fda}
\bibinfo{author}{\bibfnamefont{D.}~\bibnamefont{Hooper}},
  \bibinfo{journal}{Phys. Rev.} \textbf{\bibinfo{volume}{D91}},
  \bibinfo{pages}{035025} (\bibinfo{year}{2015}), \eprint{1411.4079}.

\bibitem[{\citenamefont{Arcadi et~al.}(2015)\citenamefont{Arcadi, Mambrini, and
  Richard}}]{Arcadi:2014lta}
\bibinfo{author}{\bibfnamefont{G.}~\bibnamefont{Arcadi}},
  \bibinfo{author}{\bibfnamefont{Y.}~\bibnamefont{Mambrini}}, \bibnamefont{and}
  \bibinfo{author}{\bibfnamefont{F.}~\bibnamefont{Richard}},
  \bibinfo{journal}{JCAP} \textbf{\bibinfo{volume}{1503}}, \bibinfo{pages}{018}
  (\bibinfo{year}{2015}), \eprint{1411.2985}.

\bibitem[{\citenamefont{Cahill-Rowley et~al.}(2015)\citenamefont{Cahill-Rowley,
  Gainer, Hewett, and Rizzo}}]{Cahill-Rowley:2014ora}
\bibinfo{author}{\bibfnamefont{M.}~\bibnamefont{Cahill-Rowley}},
  \bibinfo{author}{\bibfnamefont{J.}~\bibnamefont{Gainer}},
  \bibinfo{author}{\bibfnamefont{J.}~\bibnamefont{Hewett}}, \bibnamefont{and}
  \bibinfo{author}{\bibfnamefont{T.}~\bibnamefont{Rizzo}},
  \bibinfo{journal}{JHEP} \textbf{\bibinfo{volume}{02}}, \bibinfo{pages}{057}
  (\bibinfo{year}{2015}), \eprint{1409.1573}.

\bibitem[{\citenamefont{Ko and Tang}(2015)}]{Ko:2014loa}
\bibinfo{author}{\bibfnamefont{P.}~\bibnamefont{Ko}} \bibnamefont{and}
  \bibinfo{author}{\bibfnamefont{Y.}~\bibnamefont{Tang}},
  \bibinfo{journal}{JCAP} \textbf{\bibinfo{volume}{1501}}, \bibinfo{pages}{023}
  (\bibinfo{year}{2015}), \eprint{1407.5492}.

\bibitem[{\citenamefont{McDermott}(2014)}]{McDermott:2014rqa}
\bibinfo{author}{\bibfnamefont{S.~D.} \bibnamefont{McDermott}},
  \bibinfo{journal}{Phys. Dark Univ.} \textbf{\bibinfo{volume}{7-8}},
  \bibinfo{pages}{12} (\bibinfo{year}{2014}), \eprint{1406.6408}.

\bibitem[{\citenamefont{Kong and Park}(2014)}]{Kong:2014haa}
\bibinfo{author}{\bibfnamefont{K.}~\bibnamefont{Kong}} \bibnamefont{and}
  \bibinfo{author}{\bibfnamefont{J.-C.} \bibnamefont{Park}},
  \bibinfo{journal}{Nucl. Phys.} \textbf{\bibinfo{volume}{B888}},
  \bibinfo{pages}{154} (\bibinfo{year}{2014}), \eprint{1404.3741}.

\bibitem[{\citenamefont{Yuan and Zhang}(2014)}]{Yuan:2014rca}
\bibinfo{author}{\bibfnamefont{Q.}~\bibnamefont{Yuan}} \bibnamefont{and}
  \bibinfo{author}{\bibfnamefont{B.}~\bibnamefont{Zhang}}
  (\bibinfo{year}{2014}), \eprint{1404.2318}.

\bibitem[{\citenamefont{Petrović et~al.}(2015)\citenamefont{Petrović,
  Serpico, and Zaharijas}}]{Petrovic:2014xra}
\bibinfo{author}{\bibfnamefont{J.}~\bibnamefont{Petrović}},
  \bibinfo{author}{\bibfnamefont{P.~D.} \bibnamefont{Serpico}},
  \bibnamefont{and}
  \bibinfo{author}{\bibfnamefont{G.}~\bibnamefont{Zaharijas}},
  \bibinfo{journal}{JCAP} \textbf{\bibinfo{volume}{1502}}, \bibinfo{pages}{023}
  (\bibinfo{year}{2015}), \eprint{1411.2980}.

\bibitem[{\citenamefont{Brandt and Kocsis}(2015)}]{Brandt:2015ula}
\bibinfo{author}{\bibfnamefont{T.~D.} \bibnamefont{Brandt}} \bibnamefont{and}
  \bibinfo{author}{\bibfnamefont{B.}~\bibnamefont{Kocsis}},
  \bibinfo{journal}{Astrophys. J.} \textbf{\bibinfo{volume}{812}},
  \bibinfo{pages}{15} (\bibinfo{year}{2015}), \eprint{1507.05616}.

\bibitem[{\citenamefont{Lee et~al.}(2015)\citenamefont{Lee, Lisanti, Safdi,
  Slatyer, and Xue}}]{Lee:2015fea}
\bibinfo{author}{\bibfnamefont{S.~K.} \bibnamefont{Lee}},
  \bibinfo{author}{\bibfnamefont{M.}~\bibnamefont{Lisanti}},
  \bibinfo{author}{\bibfnamefont{B.~R.} \bibnamefont{Safdi}},
  \bibinfo{author}{\bibfnamefont{T.~R.} \bibnamefont{Slatyer}},
  \bibnamefont{and} \bibinfo{author}{\bibfnamefont{W.}~\bibnamefont{Xue}}
  (\bibinfo{year}{2015}), \eprint{1506.05124}.

\bibitem[{\citenamefont{Bartels et~al.}(2015)\citenamefont{Bartels,
  Krishnamurthy, and Weniger}}]{Bartels:2015aea}
\bibinfo{author}{\bibfnamefont{R.}~\bibnamefont{Bartels}},
  \bibinfo{author}{\bibfnamefont{S.}~\bibnamefont{Krishnamurthy}},
  \bibnamefont{and} \bibinfo{author}{\bibfnamefont{C.}~\bibnamefont{Weniger}}
  (\bibinfo{year}{2015}), \eprint{1506.05104}.

\bibitem[{\citenamefont{McDermott et~al.}(2015)\citenamefont{McDermott, Fox,
  Cholis, and Lee}}]{McDermott:2015ydv}
\bibinfo{author}{\bibfnamefont{S.~D.} \bibnamefont{McDermott}},
  \bibinfo{author}{\bibfnamefont{P.~J.} \bibnamefont{Fox}},
  \bibinfo{author}{\bibfnamefont{I.}~\bibnamefont{Cholis}}, \bibnamefont{and}
  \bibinfo{author}{\bibfnamefont{S.~K.} \bibnamefont{Lee}}
  (\bibinfo{year}{2015}), \eprint{1512.00012}.

\bibitem[{\citenamefont{Daylan et~al.}(2016)\citenamefont{Daylan, Finkbeiner,
  and Portillo}}]{probcat}
\bibinfo{author}{\bibfnamefont{T.}~\bibnamefont{Daylan}},
  \bibinfo{author}{\bibfnamefont{D.}~\bibnamefont{Finkbeiner}},
  \bibnamefont{and} \bibinfo{author}{\bibfnamefont{S.}~\bibnamefont{Portillo}},
  \emph{\bibinfo{title}{In preparation}} (\bibinfo{year}{2016}).

\bibitem[{\citenamefont{Cholis et~al.}(2015)\citenamefont{Cholis, Hooper, and
  Linden}}]{Cholis:2014lta}
\bibinfo{author}{\bibfnamefont{I.}~\bibnamefont{Cholis}},
  \bibinfo{author}{\bibfnamefont{D.}~\bibnamefont{Hooper}}, \bibnamefont{and}
  \bibinfo{author}{\bibfnamefont{T.}~\bibnamefont{Linden}},
  \bibinfo{journal}{JCAP} \textbf{\bibinfo{volume}{1506}}, \bibinfo{pages}{043}
  (\bibinfo{year}{2015}), \eprint{1407.5625}.

\bibitem[{\citenamefont{Cholis et~al.}(2014)\citenamefont{Cholis, Hooper, and
  Linden}}]{Cholis:2014noa}
\bibinfo{author}{\bibfnamefont{I.}~\bibnamefont{Cholis}},
  \bibinfo{author}{\bibfnamefont{D.}~\bibnamefont{Hooper}}, \bibnamefont{and}
  \bibinfo{author}{\bibfnamefont{T.}~\bibnamefont{Linden}}
  (\bibinfo{year}{2014}), \eprint{1407.5583}.

\bibitem[{\citenamefont{Nolan et~al.}(2012)}]{Fermi:2011bm}
\bibinfo{author}{\bibfnamefont{P.~L.} \bibnamefont{Nolan}} \bibnamefont{et~al.}
  (\bibinfo{collaboration}{Fermi-LAT Collaboration}),
  \bibinfo{journal}{Astrophys.J.Suppl.} \textbf{\bibinfo{volume}{199}},
  \bibinfo{pages}{31} (\bibinfo{year}{2012}), \eprint{1108.1435}.

\bibitem[{\citenamefont{Abdo et~al.}(2010)}]{collaboration:2010bb}
\bibinfo{author}{\bibfnamefont{A.}~\bibnamefont{Abdo}} \bibnamefont{et~al.}
  (\bibinfo{collaboration}{The Fermi-LAT collaboration})
  (\bibinfo{year}{2010}), \eprint{1003.3588}.

\bibitem[{\citenamefont{{Manchester} et~al.}(1990)\citenamefont{{Manchester},
  {Lyne}, {Johnston}, {D'Amico}, {Lim}, and {Kniffen}}}]{msps_47Tuc}
\bibinfo{author}{\bibfnamefont{R.~N.} \bibnamefont{{Manchester}}},
  \bibinfo{author}{\bibfnamefont{A.~G.} \bibnamefont{{Lyne}}},
  \bibinfo{author}{\bibfnamefont{S.}~\bibnamefont{{Johnston}}},
  \bibinfo{author}{\bibfnamefont{N.}~\bibnamefont{{D'Amico}}},
  \bibinfo{author}{\bibfnamefont{J.}~\bibnamefont{{Lim}}}, \bibnamefont{and}
  \bibinfo{author}{\bibfnamefont{D.~A.} \bibnamefont{{Kniffen}}},
  \bibinfo{journal}{Nature} \textbf{\bibinfo{volume}{345}},
  \bibinfo{pages}{598} (\bibinfo{year}{1990}).

\bibitem[{\citenamefont{{Manchester} et~al.}(1991)\citenamefont{{Manchester},
  {Lyne}, {Robinson}, {Bailes}, and {D'Amico}}}]{msps_10_47Tuc}
\bibinfo{author}{\bibfnamefont{R.~N.} \bibnamefont{{Manchester}}},
  \bibinfo{author}{\bibfnamefont{A.~G.} \bibnamefont{{Lyne}}},
  \bibinfo{author}{\bibfnamefont{C.}~\bibnamefont{{Robinson}}},
  \bibinfo{author}{\bibfnamefont{M.}~\bibnamefont{{Bailes}}}, \bibnamefont{and}
  \bibinfo{author}{\bibfnamefont{N.}~\bibnamefont{{D'Amico}}},
  \bibinfo{journal}{Nature} \textbf{\bibinfo{volume}{352}},
  \bibinfo{pages}{219} (\bibinfo{year}{1991}).

\bibitem[{\citenamefont{{Robinson} et~al.}(1995)\citenamefont{{Robinson},
  {Lyne}, {Manchester}, {Bailes}, {D'Amico}, and {Johnston}}}]{msps_more_47Tuc}
\bibinfo{author}{\bibfnamefont{C.}~\bibnamefont{{Robinson}}},
  \bibinfo{author}{\bibfnamefont{A.~G.} \bibnamefont{{Lyne}}},
  \bibinfo{author}{\bibfnamefont{R.~N.} \bibnamefont{{Manchester}}},
  \bibinfo{author}{\bibfnamefont{M.}~\bibnamefont{{Bailes}}},
  \bibinfo{author}{\bibfnamefont{N.}~\bibnamefont{{D'Amico}}},
  \bibnamefont{and}
  \bibinfo{author}{\bibfnamefont{S.}~\bibnamefont{{Johnston}}},
  \bibinfo{journal}{MNRAS} \textbf{\bibinfo{volume}{274}}, \bibinfo{pages}{547}
  (\bibinfo{year}{1995}).

\bibitem[{\citenamefont{Camilo et~al.}(2000)\citenamefont{Camilo, Lorimer,
  Freire, Lyne, and Manchester}}]{Camilo:1999fc}
\bibinfo{author}{\bibfnamefont{F.}~\bibnamefont{Camilo}},
  \bibinfo{author}{\bibfnamefont{D.}~\bibnamefont{Lorimer}},
  \bibinfo{author}{\bibfnamefont{P.~C.~C.} \bibnamefont{Freire}},
  \bibinfo{author}{\bibfnamefont{A.}~\bibnamefont{Lyne}}, \bibnamefont{and}
  \bibinfo{author}{\bibfnamefont{R.}~\bibnamefont{Manchester}},
  \bibinfo{journal}{Astrophys.J.} \textbf{\bibinfo{volume}{535}},
  \bibinfo{pages}{975} (\bibinfo{year}{2000}), \eprint{astro-ph/9911234}.

\bibitem[{\citenamefont{Bogdanov et~al.}(2006)\citenamefont{Bogdanov, Grindlay,
  Heinke, Camilo, Freire et~al.}}]{Bogdanov:2006ap}
\bibinfo{author}{\bibfnamefont{S.}~\bibnamefont{Bogdanov}},
  \bibinfo{author}{\bibfnamefont{J.~E.} \bibnamefont{Grindlay}},
  \bibinfo{author}{\bibfnamefont{C.~O.} \bibnamefont{Heinke}},
  \bibinfo{author}{\bibfnamefont{F.}~\bibnamefont{Camilo}},
  \bibinfo{author}{\bibfnamefont{P.~C.~C.} \bibnamefont{Freire}},
  \bibnamefont{et~al.}, \bibinfo{journal}{Astrophys.J.}
  \textbf{\bibinfo{volume}{646}}, \bibinfo{pages}{1104} (\bibinfo{year}{2006}),
  \eprint{astro-ph/0604318}.

\bibitem[{\citenamefont{Acero et~al.}(2015)}]{Acero:2015hja}
\bibinfo{author}{\bibfnamefont{F.}~\bibnamefont{Acero}} \bibnamefont{et~al.}
  (\bibinfo{collaboration}{Fermi-LAT}) (\bibinfo{year}{2015}),
  \eprint{1501.02003}.

\bibitem[{\citenamefont{Manchester et~al.}(2005)\citenamefont{Manchester,
  Hobbs, Teoh, and Hobbs}}]{Manchester:2004bp}
\bibinfo{author}{\bibfnamefont{R.~N.} \bibnamefont{Manchester}},
  \bibinfo{author}{\bibfnamefont{G.~B.} \bibnamefont{Hobbs}},
  \bibinfo{author}{\bibfnamefont{A.}~\bibnamefont{Teoh}}, \bibnamefont{and}
  \bibinfo{author}{\bibfnamefont{M.}~\bibnamefont{Hobbs}},
  \bibinfo{journal}{Astron.J.} \textbf{\bibinfo{volume}{129}},
  \bibinfo{pages}{1993} (\bibinfo{year}{2005}), \eprint{astro-ph/0412641}.

\bibitem[{\citenamefont{Lorimer et~al.}(2015)}]{Lorimer:2015iga}
\bibinfo{author}{\bibfnamefont{D.~R.} \bibnamefont{Lorimer}}
  \bibnamefont{et~al.}, \bibinfo{journal}{Mon. Not. Roy. Astron. Soc.}
  \textbf{\bibinfo{volume}{450}}, \bibinfo{pages}{2185} (\bibinfo{year}{2015}),
  \eprint{1501.05516}.

\bibitem[{\citenamefont{{Arons}}(1996)}]{1996A&AS..120C..49A}
\bibinfo{author}{\bibfnamefont{J.}~\bibnamefont{{Arons}}},
  \bibinfo{journal}{\aaps} \textbf{\bibinfo{volume}{120}}, \bibinfo{pages}{C49}
  (\bibinfo{year}{1996}).

\bibitem[{\citenamefont{{Ruderman} and
  {Sutherland}}(1975)}]{1975ApJ...196...51R}
\bibinfo{author}{\bibfnamefont{M.~A.} \bibnamefont{{Ruderman}}}
  \bibnamefont{and} \bibinfo{author}{\bibfnamefont{P.~G.}
  \bibnamefont{{Sutherland}}}, \bibinfo{journal}{\apj}
  \textbf{\bibinfo{volume}{196}}, \bibinfo{pages}{51} (\bibinfo{year}{1975}).

\bibitem[{\citenamefont{{Harding}}(1981)}]{1981ApJ...245..267H}
\bibinfo{author}{\bibfnamefont{A.~K.} \bibnamefont{{Harding}}},
  \bibinfo{journal}{\apj} \textbf{\bibinfo{volume}{245}}, \bibinfo{pages}{267}
  (\bibinfo{year}{1981}).

\bibitem[{\citenamefont{{Faucher-Gigu{\`e}re} and
  {Kaspi}}(2006)}]{2006ApJ...643..332F}
\bibinfo{author}{\bibfnamefont{C.-A.} \bibnamefont{{Faucher-Gigu{\`e}re}}}
  \bibnamefont{and} \bibinfo{author}{\bibfnamefont{V.~M.}
  \bibnamefont{{Kaspi}}}, \bibinfo{journal}{\apj}
  \textbf{\bibinfo{volume}{643}}, \bibinfo{pages}{332} (\bibinfo{year}{2006}),
  \eprint{astro-ph/0512585}.

\bibitem[{\citenamefont{Linden}(2015)}]{Linden:2015qha}
\bibinfo{author}{\bibfnamefont{T.}~\bibnamefont{Linden}}
  (\bibinfo{year}{2015}), \eprint{1509.02928}.

\end{thebibliography}
\bibliographystyle{apsrev}

\end{document}